\documentclass[aps,pre,onecolumn,floatfix,nofootinbib]{revtex4}

\usepackage{amssymb,amsfonts,amsmath}
\usepackage{graphicx}
\usepackage{dcolumn}
\usepackage{bm}
\usepackage{mathrsfs}
\usepackage{mhchem}
\usepackage{subfigure}
\usepackage{color}

\usepackage{booktabs}
\usepackage{epstopdf}
\usepackage{multirow}

\usepackage{chemarrow}
\usepackage{extarrows}
\usepackage{mathtools}

\usepackage{epsf}
\usepackage{epstopdf}
\newcommand{\be}{\begin{equation}}
\newcommand{\ee}{\end{equation}}
\newcommand{\bea}{\begin{eqnarray}}
\newcommand{\eea}{\end{eqnarray}}

\usepackage[T3,T1]{fontenc}
\DeclareSymbolFont{tipa}{T3}{cmr}{m}{n}
\DeclareMathAccent{\invbreve}{\mathalpha}{tipa}{16}

\newcommand{\bUps}{\boldsymbol{\Upsilon}}
\newcommand{\bPsi}{\boldsymbol{\Psi}}
\newcommand{\bPhi}{\boldsymbol{\Phi}}

\begin{document}

\title{The stress tensor of dilute active colloidal suspensions}

\author{Pierre Gaspard}
\thanks{ORCID: {\tt 0000-0003-3804-2110}}
\affiliation{ Center for Nonlinear Phenomena and Complex Systems, Universit{\'e} Libre de Bruxelles (U.L.B.), Code Postal 231, Campus Plaine, B-1050 Brussels, Belgium}


\begin{abstract}
The stress tensor is calculated for dilute active suspensions composed of colloidal Janus particles propelled by self-diffusiophoresis and powered by a chemical reaction.  The Janus particles are assumed to be spherical and made of catalytic and non-catalytic hemispheres.  The chemical reaction taking place on the catalytic part of each Janus particle generates local molecular concentration gradients at the surface of the particle and, thus, an interfacial velocity slippage between the fluid and the solid particle, which is the propulsion mechanism of self-diffusiophoresis.  In the dilute-system limit, the contributions of the suspended particles to the stress tensor are calculated by solving the chemohydrodynamic equations for the fluid velocity and the molecular concentrations around every Janus particle considered as isolated and far apart from each other.  The results are the following.  First, the well-known Einstein formula for the effective shear viscosity of colloidal suspensions is recovered, including the effect of a possible Navier slip length.  Next, two further contributions are obtained, which depend on the molecular concentrations of the fuel and product species of the chemical reaction and on the orientation of the Janus particles.  The second contribution is caused by simple diffusiophoresis, which already exists in passive suspensions with global concentration gradients and no reaction.  The third contribution is due to the self-diffusiophoresis generated by the chemical reaction, which arises in active suspensions.  The calculation gives quantitative predictions, which depend on the constitutive properties of the fluid and the fluid-solid interfaces, as well as on the geometry of the Janus particles.
\end{abstract}


\maketitle

\section{Introduction}
\label{sec:intro}

Active colloidal suspensions are non-equilibrium systems composed of solid particles with sizes significantly larger than molecular diameters and moving in a fluid by harnessing energy from their environment.  Like in other kinds of active matter, the propulsion of these particles results from the conversion of the energy supplied by the surroundings into mechanical motion \cite{PKOSACMLC04,FAMO05,WDAMS13,SSK14,MJRLPRA13,BDLRVV16}.  In an important family of such systems, the solid particles are partly coated with some catalytic material, where the fuel coming from the surrounding fluid can react by heterogeneous catalysis and generate local concentration gradients, driving the particle by self-diffusiophoresis or self-electrophoresis \cite{A89,AP91,W15,OPD17,IGL17,MP17,KKH18}.  

Examples of such active particles are silica spheres with a platinum cap and moving with speeds of the order of $10~\mu$m/s in an aqueous solution of hydrogen peroxide \cite{VTZKGKO10,KYCS10}.  In another example, hydrazine is used as fuel and iridium as catalyst, leading to higher speeds of about $20~\mu$m/s \cite{GPDW14}.  Since the particles are made of two parts, namely, the catalytic and the non-catalytic ones, they are called Janus particles.

In such examples, the propulsion mechanism is based on diffusiophoresis, in which a velocity slippage is generated between the fluid and the solid particle, as a consequence of local molecular concentration gradients along the fluid-solid interface and, thus, gradients in the interaction forces between the molecules of the fluid and the atoms composing the different parts of the particle surface \cite{GLA05,GLA07,K13,CRRK14}.  The so-generated interfacial velocity slippage induces a local flow of the fluid around the particle, which leads to its propulsion \cite{RHSK17,CEIG19}.

These effects can be described in terms of the constitutive properties of the aqueous solution and the fluid-solid interfaces, combined with the geometry of the Janus particles.  The aqueous solution is an incompressible fluid characterised by its shear viscosity and the diffusion coefficients of the fuel and product molecules.  The constitutive properties of the fluid-solid interfaces include the rate constants of the catalytic surface reactions, the diffusiophoretic constants, and the possible Navier slip lengths.  Since Janus particles have sizes ranging from hundreds of nanometers to a few micrometers, there is a large separation of length scales with respect to the diameters of the atoms and molecules composing the system, so that the fluid can be described in terms of the Navier-Stokes equations of hydrodynamics coupled by interfacial reactions and diffusiophoresis to the advection-diffusion equations for the molecular concentrations, while the active colloids can be described as rigid Janus particles.  In this way, the macroscopic coarse-grained equations ruling the active suspension as a continuous medium can be deduced from the mesoscopic description based on the chemohydrodynamics of the fluid containing the colloidal particles.  Such an approach allows us to predict the macroscopic properties of active matter and to justify the coarse-grained theories from the underlying structure and dynamics of its components.

This approach has already been developed for dilute active suspensions, where the particles move in a fluid that is globally at rest \cite{GK20}.  In dilute suspensions, the particles are largely separated from each other, so that their mutual interactions can be neglected and they can be considered as isolated in the fluid.  As a consequence, the macroscopic diffusion-reaction equations ruling the colloidal and molecular species can be deduced together with the complete expression for the entropy production rate density of such dilute active suspensions, as obtained in Ref.~\cite{GK20}.  

However, the extension of these results to active suspensions with a fluid flowing on large scales, requires the knowledge of the stress tensor, including its active part \cite{KJJPS05,R10,MJRLPRA13,JGS18,CT18}.  In principle, the stress tensor can be deduced from the underlying fluid mechanics of the suspension, using methods pioneered in 1906 by Einstein in the case of passive rather than active systems \cite{E1906,LL59,B70,H77,BKM77,BKS06}.

\begin{figure}[h]
\centerline{\scalebox{0.5}{\includegraphics{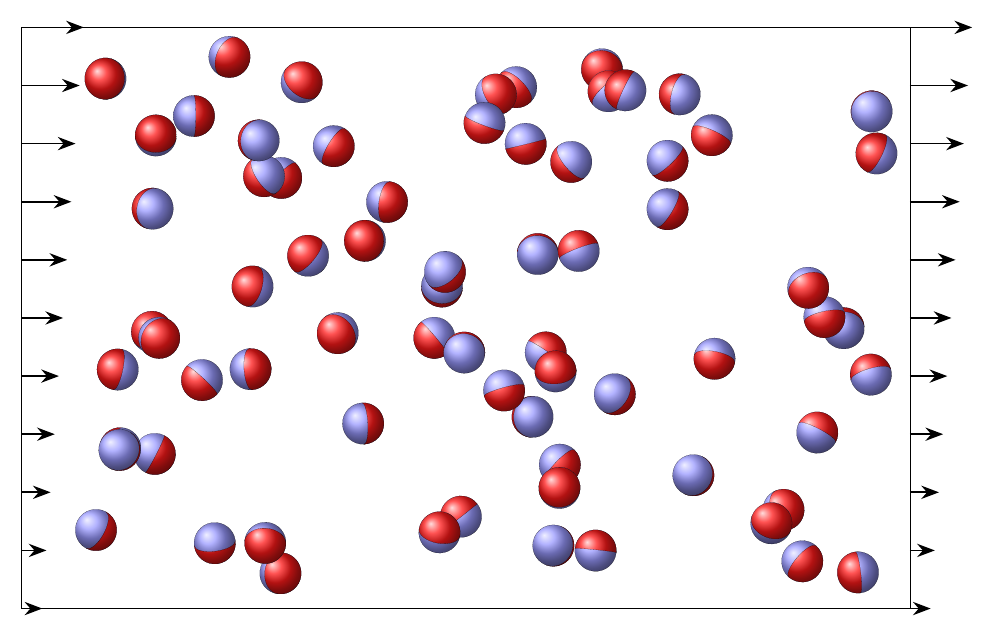}}}
\caption{Schematic representation of an active suspension of Janus particles immersed in a shear flow.}
\label{fig1}
\end{figure}

The purpose of this paper is to carry out this programme for a dilute suspension of axisymmetric Janus particles immersed in an aqueous solution containing fuel and product species, which react on a catalytic hemisphere of the particles. A schematic representation of the system of interest is shown in Fig.~\ref{fig1}.

The study of this system is important not only because it has experimental realisations, but also because the mathematical deduction of the active stress tensor in dilute suspensions made of synthetic Janus particles can provide guiding lines to infer the active stress tensor in more complex forms of active matter such as active gels, suspensions of bacteria or other unicellular organisms, and  biological tissues like muscles.

Our goal is thus to deduce the stress tensor of the active suspension of Janus particles under dilute conditions, following the approach developed in Ref.~\cite{GK20}.  The novelty of this approach is that the description is based on the distribution function of the positions and the orientations of the Janus particles, providing a clear understanding of the ways the couplings between the fluid flow, the diffusion of molecular species, and their reactions may depend on the orientation of the particles.  Afterwards, the macroscopic theories formulated in terms of the macrofields of polar or nematic ordering \cite{KJJPS05,R10,MJRLPRA13,JGS18,CT18} can be deduced therefrom.  The mesoscopic and macroscopic levels of description of active suspensions can thus be bridged together within the framework of this approach.

After this introduction, the plan of the paper is the following.  The chemohydrodynamics of dilute active suspensions is presented in Sec.~\ref{sec:chemohydro}.  The stationary fields around a single isolated Janus particle are obtained in Sec.~\ref{sec:one-body_fields}.  The details of the calculations for the one-particle profiles of molecular concentrations are given in App.~\ref{app:concentrations}.  After a summary of the definitions and relevant properties of the vectorial spherical harmonics in App.~\ref{app:VSH}, the calculations of the velocity field around a single isolated Janus particle are detailed in App.~\ref{app:velocity}.  The stress tensor of the dilute active suspension is deduced in Sec.~\ref{sec:stress_tensor}, using the calculations of App.~\ref{app:P_ij}.  Section~\ref{sec:uniform-case} is devoted to the case of Janus particles with uniform diffusiophoretic constants.  The results obtained in Ref.~\cite{GK20} for the translational and rotational velocities are adapted in App.~\ref{app:velocities-uniform} to the case of the Janus particles that are discussed in Sec.~\ref{sec:uniform-case}.  Conclusion and perspectives are drawn in Sec.~\ref{sec:conclusion}.

\section{The chemohydrodynamics of dilute active suspensions}
\label{sec:chemohydro}

For the purpose of deriving the stress tensor of a dilute suspension of active Janus particles, the mathematical formulation is here set up on the basis of chemohydrodynamics, which rules the motion of the fluid containing force-free and torque-free particles.

\subsection{Description of the system}

The suspension is composed of many spherical Janus particles of micrometric radius $R$ immersed in an incompressible aqueous solution containing the molecular species ${\rm A}$ and ${\rm B}$.  The Janus particles have catalytic and non-catalytic hemispheres, which are denoted ${\rm c}$ and ${\rm n}$, respectively.  The reversible reaction
\be
{\rm A} + {\rm C} \underset{\kappa_-^{\rm c}}{\stackrel{\kappa_+^{\rm c}}{\rightleftharpoons}}  {\rm B} + {\rm C}
\label{reaction}
\ee
happens on the catalytic hemisphere with rate constants per unit area $\kappa_\pm^{\rm c}$, ${\rm C}$ denoting the colloidal particles as third solute species in addition to the molecular species ${\rm A}$ and ${\rm B}$.  The Janus particles are propelled by self-diffusiophoresis caused by the local concentration profiles of ${\rm A}$ and ${\rm B}$ generated around each particle by the reaction~(\ref{reaction}).  There is no reaction in the bulk of the solution.

The suspension is described in terms of the velocity ${\bf v}({\bf r},t)$ of the fluid and the molecular concentrations $c_{\rm A}({\bf r},t)$ and $c_{\rm B}({\bf r},t)$, which are fields defined in space ${\bf r}=(x,y,z)$ and time $t$.  In addition, the suspension contains $N_{\rm C}$ colloidal particles, which are labelled with the index $c=1,2,\dots,N_{\rm C}$.  Each Janus particle has a position ${\bf R}_c$, an orientation given by the unit vector ${\bf u}_c$ with direction along the symmetry axis and pointing towards its catalytic hemisphere, a velocity ${\bf V}_c$, and an angular velocity $\boldsymbol{\Omega}_c$.  The orientation of each particle is equivalently described by the polar and azimuthal angles $(\theta_c,\varphi_c)$, such that ${\bf u}_c=\sin\theta_c \cos\varphi_c \, \boldsymbol{1}_x+\sin\theta_c \sin\varphi_c \, \boldsymbol{1}_y+\cos\theta_c \, \boldsymbol{1}_z$, where $\{\boldsymbol{1}_x, \boldsymbol{1}_y,  \boldsymbol{1}_z\}$ is a Cartesian basis of orthogonal unit vectors of the tridimensional space.
The velocity field inside each Janus particle is thus given by
\be
{\bf v}_{\rm solid}^{(c)}({\bf r},t) = {\bf V}_c(t) + \boldsymbol{\Omega}_c(t) \times \left[ {\bf r}-{\bf R}_c(t)\right]
\qquad\mbox{for}\qquad
\Vert{\bf r}-{\bf R}_c(t)\Vert < R \, .
\label{velocity-solid}
\ee

\subsection{The equations for the velocity field of the fluid}

Since the fluid is incompressible and the colloidal particles are rigid, the velocity field of the suspension satisfies the condition
\be
\boldsymbol{\nabla}\cdot{\bf v} = 0 \, .
\label{div.v=0}
\ee
Furthermore, the velocity field obeys the Navier-Stokes equations
\be
\rho \left( \partial_t + {\bf v}\cdot\boldsymbol{\nabla}\right) {\bf v} = \boldsymbol{\nabla}\cdot{\boldsymbol{\sigma}} \, ,
\label{NS-eqs}
\ee
where $\rho$ is the mass density of the fluid and ${\boldsymbol{\sigma}}=(\sigma_{ij})$ is the stress tensor.  The latter is related to the pressure tensor $\boldsymbol{\mathsf P}=(P_{ij})$ by ${\boldsymbol{\sigma}}=-{\boldsymbol{\mathsf P}}$.  The components of the stress tensor are given by
\be
\sigma_{ij} = -p \, \delta_{ij} +\eta_0 \left(\nabla_i v_j+\nabla_j v_i\right) + \sum_{c=1}^{N_{\rm C}} \sigma_{ij}^{(c)} \, ,
\label{stress-tensor-susp}
\ee
$p$ being the pressure, $\eta_0$ the shear viscosity of the fluid, and $\sigma_{ij}^{(c)}$ the contribution of the $c^{\footnotesize\mbox{th}}$ colloidal particle to the stress tensor.  The latter can be expressed as a distribution that is singular at the interface between the particle and the fluid and that is determined by the boundary conditions on the velocity field \cite{MB74,ABM75,GK18,GK19}.  These boundary conditions include the effects of sliding friction due to possible partial slip between the fluid and the particle, and by diffusiophoresis coupling the fluid velocity with the concentrations of the molecular species.  The surface $\Vert{\bf r}-{\bf R}_c(t)\Vert=R$ of the $c^{\footnotesize\mbox{th}}$ particle is denoted $\Sigma_c(t)$.  The unit vector normal to this surface and oriented towards the fluid is denoted ${\bf n}_c=R^{-1}({\bf r}-{\bf R}_c)$ for ${\bf r}\in\Sigma_c(t)$.  In the direction normal to the interface, the boundary condition reads
\be
{\bf n}_c\cdot\left({\bf v} -{\bf v}_{\rm solid}^{(c)}\right)_{\Sigma_c} = 0\, .
\label{bc-v-1}
\ee
In the two directions tangential to the interface, they are given by
\be
\left({\boldsymbol{\mathsf 1}}-{\bf n}_c{\bf n}_c\right) \cdot \left[{\bf v} -{\bf v}_{\rm solid}^{(c)}
- b\left(\pmb{\nabla}{\bf v}+\pmb{\nabla}{\bf v}^{\rm T}\right)\cdot{\bf n}_c +\sum_\varkappa b_\varkappa \pmb{\nabla}c_\varkappa\right]_{\Sigma_c} = 0 \, , 
\label{bc-v-2}
\ee
where ${\boldsymbol{\mathsf 1}}=(\delta_{ij})$ is the tridimensional identity matrix, $b$ is the Navier slip length characterising the possible partial slip between the particle and the fluid, and $b_\varkappa$ is the diffusiophoretic constant of the molecular species $\varkappa\in\{{\rm A},{\rm B}\}$.  $(\pmb{\nabla}{\bf v})^{\rm T}$ denotes the transpose of the matrix $\pmb{\nabla}{\bf v}$.

The Navier slip length $b$ is assumed to be uniform on the whole interface between the spherical particle and the fluid.  The diffusiophoretic constants are supposed to take different values on each hemisphere
\be
b_\varkappa = \sum_{h={\rm c},{\rm n}} b_\varkappa^h \, H^h(\theta_c) = b_\varkappa^{\rm c} \, H^{\rm c}(\theta_c) + b_\varkappa^{\rm n} \, H^{\rm n}(\theta_c) \, , 
\label{diffusio-coeff}
\ee
where $H^{h}(\theta_c)$ denotes the Heaviside function that is equal to one on hemisphere $h$ and zero otherwise.  The catalytic hemisphere is taken as $0\le\theta_c\le\frac{\pi}{2}$ and the non-catalytic hemisphere as $\frac{\pi}{2}<\theta_c\le\pi$ in the frame where the polar axis is oriented in the direction of the unit vector ${\bf u}_c$.

\subsection{The equations for the molecular concentrations}

The concentrations of the molecular species $\varkappa\in\{{\rm A},{\rm B}\}$ satisfy the following advection-diffusion-reaction equations
\be
\left(\partial_t  + {\bf v} \cdot \pmb{\nabla}\right) c_\varkappa = D_\varkappa {\nabla}^2 c_\varkappa + \sum_{c=1}^{N_{\rm C}} \nu_\varkappa \, w^{(c)} \, , 
\label{ADR-eqs}
\ee
where $D_\varkappa$ is the associated molecular diffusivity, $\nu_\varkappa$ is the stoichiometric coefficient of species $\varkappa$ in the reaction ${\rm A}\to{\rm B}$ of Eq.~(\ref{reaction}) ($\nu_{\rm A}=-1$ and $\nu_{\rm B}=+1$), and $w^{(c)}$ is the contribution of the $c^{\footnotesize\mbox{th}}$ particle to the reaction rate density.\footnote{If several reactions occurred, $\nu_\varkappa \, w^{(c)}$ should be replaced by $\sum_r \nu_{\varkappa r} \, w_r^{(c)}$, where $\nu_{\varkappa r}$ is the stoichiometric coefficient of species~$\varkappa$ in reaction~$r$ and $w_r^{(c)}$ is the corresponding rate density.}  This contribution is also singular at the interface between the Janus particle and the fluid.  It is fixed by the boundary conditions
\be
D_\varkappa \left( {\bf n}_c\cdot\pmb{\nabla} c_\varkappa \right)_{\Sigma_c} = -\nu_\varkappa \, (\kappa_+\, c_{\rm A}-\kappa_-\, c_{\rm B})_{\Sigma_c} \, , 
\label{bc-ck}
\ee
where 
\be
\kappa_{\pm} = \kappa_{\pm}^{\rm c} \, H^{\rm c}(\theta_c)
\label{kappa}
\ee
are the rate constants per unit area of the forward and reverse surface reactions~(\ref{reaction}).

The overall reaction rate of the $c^{\footnotesize\mbox{th}}$ Janus particle is defined as the number of reactive events ${\rm A}\to{\rm B}$ per unit time taking place on the catalytic hemisphere of the particle, A being considered as the fuel and B as the product.  This overall reaction rate is given by the incoming flux of fuel molecules A (or the outgoing flux of product molecules B) at the surface of the Janus particle,
\be\label{overall_reaction_rate}
W_c \equiv \int_{\Sigma_c} D_{\rm A}\, \boldsymbol{\nabla} c_{\rm A}\cdot d\boldsymbol{\Sigma} = - \int_{\Sigma_c} D_{\rm B}\, \boldsymbol{\nabla} c_{\rm B}\cdot d\boldsymbol{\Sigma} = \int_{\Sigma_c} (\kappa_+\, c_{\rm A}-\kappa_-\, c_{\rm B}) \, d\Sigma \, ,
\ee
as expressed in terms of the surface area element $d{\boldsymbol{\Sigma}}={\bf n}_c d\Sigma$ and the rate constants~(\ref{kappa}).

We note that the suspension is active if the molecular concentrations are not in their Guldberg-Waage equilibrium ratio, i.e., if $\kappa_+^{\rm c}\, c_{\rm A} \ne \kappa_-^{\rm c} \, c_{\rm B}$.  Otherwise, the suspension is said to be passive.  Even if the suspension is at chemical equilibrium, i.e., if $\kappa_+^{\rm c}\, c_{\rm A} = \kappa_-^{\rm c} \, c_{\rm B}$, the system may evolve out of equilibrium under the effects of non-zero gradients of velocity or molecular concentrations.
The suspension is thus active because the chemical reaction is out of equilibrium.

\subsection{The force and the torque exerted by the fluid on each particle}

Once the velocity field of the fluid is known, the force and the torque exerted by the fluid on each colloidal particle are respectively given by
\bea
{\bf F}_c &=& \int_{\Sigma_c} \boldsymbol{\sigma}^{({\rm fl})}\cdot d{\boldsymbol{\Sigma}} \, ,\label{force}\\
{\bf T}_c &=& \int_{\Sigma_c} ({\bf r}-{\bf R}_c)\times \left(\boldsymbol{\sigma}^{({\rm fl})}\cdot d{\boldsymbol{\Sigma}}\right) ,\label{torque}
\eea
where $\boldsymbol{\sigma}^{({\rm fl})}=-p\, \boldsymbol{\mathsf 1}+\eta_0(\boldsymbol{\nabla}{\bf v}+\boldsymbol{\nabla}{\bf v}^{\rm T})$ is the stress tensor of the fluid around the particle and $c=1,2,\dots,N_{\rm C}$.

\subsection{The distribution function of the Janus particles}

The distribution of the positions and orientations of the Janus particles is defined by
\be
\invbreve{f}_{\rm C}({\bf r},{\bf u},t) \equiv \sum_{c=1}^{N_{\rm C}} \delta^3[{\bf r}-{\bf R}_c(t)] \, \delta^2[{\bf u}-{\bf u}_c(t)] \, ,
\label{f-dfn}
\ee
where $\delta^3[{\bf r}-{\bf R}_c(t)]$ is the tridimensional Dirac distribution centered at the position ${\bf R}_c$ of the $c^{\footnotesize\mbox{th}}$ particle and $\delta^2[{\bf u}-{\bf u}_c(t)]$ stands for $\delta(\theta-\theta_c)\delta(\varphi-\varphi_c)/\sin\theta_c$.  The distribution~(\ref{f-dfn}) is normalised to the invariant total number of particles according to $N_{\rm C} = \int \invbreve{f}_{\rm C}({\bf r},{\bf u},t)\, d^3r \, d^2u$ with $d^2u = d\Omega=\sin\theta d\theta d\varphi$.

The suspension may be described as a continuous medium over length scales larger than the size of coarse-graining cells ${\cal V}_{\bf r}$ of volume $V_{\bf r}$ and centered around the positions $\bf r$, if the statistical properties of the suspension may be assumed to remain quasi-uniform in these cells.  The cells ${\cal V}_{\bf r}$ should contain a sufficiently large number of particles to justify a statistical formulation.  Accordingly, we may introduce the spatial averages of the quantities of interest over the cells ${\cal V}_{\bf r}$ as
\be
\overline{X({\bf r},t)} \equiv \frac{1}{V_{\bf r}} \int_{{\cal V}_{\bf r}} X({\bf r},t) \, d^3r \, .
\label{space-average}
\ee
In particular, the mean distribution function of the particles may thus be evaluated as
\be
f_{\rm C}({\bf r},{\bf u},t) = \overline{\invbreve f_{\rm C}({\bf r},{\bf u},t)} \equiv \frac{1}{V_{\bf r}} \int_{{\cal V}_{\bf r}} \invbreve f_{\rm C}({\bf r},{\bf u},t) \, d^3r \, ,
\label{mean-f}
\ee
such that
\be
\frac{1}{V_{\bf r}} \sum_{c:\,{\bf R}_c\in{\cal V}_{\bf r}} (\cdot) = \int d^2u \, f_{\rm C}({\bf r},{\bf u},t) \, (\cdot) 
\label{coarse-graining}
\ee
for any quantity $(\cdot)$ depending on the coordinates $\{{\bf R}_c,{\bf u}_c\}_{c=1}^{N_{\rm C}}$ of the Janus particles.  
We note that we may also consider a statistical ensemble of configurations for the Janus particles and the corresponding statistical average $\langle\cdot\rangle$, which should be equal to the spatial average, $f_{\rm C}({\bf r},{\bf u},t)=\langle \invbreve f_{\rm C}({\bf r},{\bf u},t)\rangle$, if ergodicity held \cite{B70}.

The density of colloidal particles is given by
\be
n_{\rm C}({\bf r},t) \equiv \int d^2u \, f_{\rm C}({\bf r},{\bf u},t) \, .
\label{n_C}
\ee
In addition, we may introduce the moments of the distribution of the orientations of the Janus particles as
\bea
&& p_i({\bf r},t) \equiv \int d^2u \, f_{\rm C}({\bf r},{\bf u},t) \, u_i \, , \label{p_i}\\
&& q_{ij}({\bf r},t) \equiv \int d^2u \, f_{\rm C}({\bf r},{\bf u},t) \, Q_{ij}({\bf u})
\qquad\mbox{with}\qquad
Q_{ij}({\bf u}) \equiv u_i \, u_j - \frac{1}{3} \, \delta_{ij} \, , \label{q_ij}\\
&& r_{ijk}({\bf r},t) \equiv \int d^2u \, f_{\rm C}({\bf r},{\bf u},t) \, R_{ijk}({\bf u})
\qquad\mbox{with}\qquad
R_{ijk}({\bf u}) \equiv u_i \, u_j \, u_k - \frac{1}{5} \, u_i \, \delta_{jk} - \frac{1}{5} \, u_j \, \delta_{ik} - \frac{1}{5} \, u_k \, \delta_{ij}\, , \label{r_ijk}\\
&&\qquad\qquad\qquad\qquad \vdots\nonumber
\eea
such that the distribution function can be decomposed as
\be
f_{\rm C}({\bf r},{\bf u},t) = \frac{1}{4\pi} \, n_{\rm C}({\bf r},t) + \frac{3}{4\pi} \, p_i({\bf r},t) \, u_i 
+ \frac{15}{8\pi} \, q_{ij}({\bf r},t) \, Q_{ij}({\bf u}) + \frac{35}{8\pi} \, r_{ijk}({\bf r},t) \, R_{ijk}({\bf u}) + \cdots
\label{f_C-decomp}
\ee
with the convention of summation over repeated indices.  We note that the tensors $q_{ij}$ and $Q_{ij}$ are symmetric and traceless, while the tensors $r_{ijk}$ and $R_{ijk}$ are totally symmetric in the indices and traceless in two indices, i.e., satisfying $r_{iik}=0$ and $R_{iik}=0$, as well as similar identities obtained by the permutation of indices.

\subsection{Dilute suspensions}

The suspension is dilute if the mean distance ${\cal R}_{\rm C}$ between nearest neighbouring particles is much larger than their radius $R$:
\be
{\cal R}_{\rm C} \sim n_{\rm C}^{-1/3} \gg R \, .
\label{dilute-susp}
\ee
In this case, the interactions between the colloidal particles can be neglected and the suspension behaves as a collection of largely separated and quasi-isolated particles immersed in the coarse-grained velocity and concentration fields.

Since the cells ${\cal V}_{\bf r}$ used for coarse graining contain many colloidal particles, they can be decomposed as ${\cal V}_{\bf r}=\bigcup_{c:\,{\bf R}_c\in{\cal V}_{\bf r}}{\cal V}_{\bf r}^{(c)}$ into smaller domains ${\cal V}_{\bf r}^{(c)}$, each containing a single particle, for instance using a Voronoi decomposition.  The size of these domains is of the order of the mean distance between nearest neighbours, ${\cal R}_{\rm C}\sim n_{\rm C}^{-1/3}$, satisfying the condition~(\ref{dilute-susp}) in a dilute suspension.   As a consequence, the rim of each domain ${\cal V}_{\bf r}^{(c)}$ is at large distance from the Janus particle located at its center.  Inside each domain ${\cal V}_{\bf r}^{(c)}$, the profiles of the velocity, pressure, and concentration fields are expected to take the following forms,
\bea
&& {\bf v} = {\bf v}^{\circ} + {\boldsymbol{\mathsf G}}\cdot{\bf r} + \Delta{\bf v} \, , \label{asympt-v}\\
&& p = p^{\circ} + \Delta p \, , \label{asympt-p}\\
&& c_\varkappa = c_\varkappa^{\circ} + {\bf g}_\varkappa \cdot {\bf r} + \Delta c_\varkappa 
\qquad\mbox{for}\qquad \varkappa\in\{{\rm A},{\rm B}\} \, ,
\label{asympt-ck}
\eea
as expressed in a frame centered on the position of each particle, i.e., such that ${\bf R}_c=0$.
In Eq.~(\ref{asympt-v}), ${\boldsymbol{\mathsf G}}=(G_{ij})$ denotes a possible tensor of velocity gradients holding on average across the domain ${\cal V}_{\bf r}^{(c)}$.  Since the fluid is incompressible, this tensor is traceless, ${\rm tr}\,{\boldsymbol{\mathsf G}}=G_{ii}=0$.  Moreover, ${\bf v}^{\circ}$ is a constant value for the fluid velocity and $\Delta{\bf v}$ is the disturbance velocity due to the presence of the Janus particle in the center of the domain ${\cal V}_{\bf r}^{(c)}$.  This disturbance velocity vanishes at large distances from the particle. Similarly, in Eqs.~(\ref{asympt-p}) and~(\ref{asympt-ck}), $p^{\circ}$ and $c_\varkappa^{\circ}$ denote constant values, and $\Delta p$ and $\Delta c_\varkappa$ the disturbance pressure and concentrations, vanishing away from the particle.  In Eq.~(\ref{asympt-ck}), ${\bf g}_\varkappa$ denotes a possible gradient in the profile of the molecular concentration $c_\varkappa$, holding on average across the domain ${\cal V}_{\bf r}^{(c)}$.

\subsection{Linearization of the equations of motion and stationarity}

We suppose that the flow is laminar with a low Reynolds number
\be
{\rm Re} \equiv \frac{\rho V_c R}{\eta_0} \ll 1 \, ,
\label{Re}
\ee
$V_c\equiv \Vert {\bf V}_c\Vert$ being the speed of the particles; and low P\'eclet numbers
\be
{\rm Pe}_\varkappa \equiv \frac{V_c R}{D_\varkappa} \ll 1 \, .
\label{Pe}
\ee
These conditions are satisfied in aqueous suspensions, where the active particles are micrometric ($R\sim 1 \,\mu$m) and have speeds of the order of $V_c \sim 10\, \mu$m/s, so that ${\rm Re}\sim 10^{-5}$ for $\eta_0/\rho\simeq 10^{-6}\,$m$^2$/s, and ${\rm Pe}_\varkappa\sim 10^{-2}$ for $D_\varkappa \sim 10^{-9}\,$m$^2$/s.  Accordingly, the Navier-Stokes equations~(\ref{NS-eqs}) reduce to the linear Stokes equations and the advection-diffusion-reaction equations~(\ref{ADR-eqs}) can also be linearized by neglecting ${\bf v}\cdot\boldsymbol{\nabla}(\cdot)$ in front of $\partial_t(\cdot)$.

Furthermore, the velocity and concentration fields are assumed to vary over time scales that are longer than those associated with viscosity and molecular diffusion, $t_{\rm visc}\sim \rho R^2/\eta_0$ and $t_{\rm diff}\sim R^2/D_\varkappa$, respectively.  As a consequence, we may consider stationary solutions for the linearized equations~(\ref{NS-eqs}) and~(\ref{ADR-eqs}) in order to determine the profiles of the fields around each Janus particle.  Therefore, these equations reduce to
\bea
\eta_0 \nabla^2 {\bf v} &=& \boldsymbol{\nabla} p \, , \label{Stokes-eqs} \\
\nabla^2 c_\varkappa &=& 0 \, , \label{Laplace-eq-ck}
\eea
which should be solved with the boundary conditions~(\ref{bc-v-1}), (\ref{bc-v-2}), and (\ref{bc-ck}) for a single colloidal particle located at ${\bf r}=0$, and
\bea
&&(\nabla_i v_j)_\infty = G_{ji} \, , \label{bc-v-inf}\\
&&(\boldsymbol{\nabla} c_\varkappa)_\infty = {\bf g}_\varkappa
\qquad\mbox{for}\qquad \varkappa\in\{{\rm A},{\rm B}\} \, ,
\label{bc-ck-inf}
\eea
at large distances from the particle, i.e., for $\Vert{\bf r}\Vert \gg R$.

\subsection{The effective stress tensor of the suspension}

We suppose that the particles are force free, i.e., the forces~(\ref{force}) are equal to zero, ${\bf F}_c=0$.  Under these conditions, the disturbance velocity and pressure induced by each particle behave as
\bea
&& \Delta{\bf v} = O\Big(\frac{1}{r^2}\Big) \, , \label{Dv-O}\\
&& \Delta p = O\Big(\frac{1}{r^3}\Big) \, , \label{Dp-O}
\eea
for $r=\Vert{\bf r}\Vert \to\infty$.  In such dilute suspensions, the effective stress tensor can be calculated as follows using the spatial average~(\ref{space-average}),
\be
\overline{\sigma}_{ij} = -\overline{p}\, \delta_{ij} + \eta_0\, \overline{\nabla_i v_j+\nabla_j v_i} + \frac{1}{V_{\bf r}} \sum_{c:\,{\bf R}_c\in{\cal V}_{\bf r}} \int_{{\cal V}_{\bf r}^{(c)}} \left[ \Delta \sigma_{ik}^{({\rm fl})} \, r_j \, d\Sigma_k - \eta_0 \left( \Delta v_i \, d\Sigma_j + \Delta v_j \, d\Sigma_i \right)\right] ,
\label{Pij-suspension-LL}
\ee
where $\Delta \sigma_{ik}^{({\rm fl})} = -\Delta p \, \delta_{ik} +\eta_0 \left( \nabla_i \Delta v_k + \nabla_k \Delta v_i\right)$ is the disturbance in the stress tensor of the fluid due to the presence of a particle at the center of the cell ${\cal V}_{\bf r}^{(c)}$ and evaluated at large distances from this center, after having solved the one-particle problem \cite{LL59,B70}. In Eq.~(\ref{Pij-suspension-LL}), the spatial averages of the velocity gradients correspond to the velocity gradients in the domains ${\cal V}_{\bf r}^{(c)}$ as introduced in Eq.~(\ref{asympt-v}) according to $\overline{\nabla_i v_j}=G_{ji}$ and $\overline{\nabla_j v_i}=G_{ij}$.

Furthermore, the particles are assumed to be torque free, i.e., the torques~(\ref{torque}) are equal to zero, ${\bf T}_c=0$.  Consequently, the stress tensor of the suspension remains symmetric, $\sigma_{ij}=\sigma_{ji}$, as shown below.  

The calculation of the stress tensor will be carried out for the dilute active suspension in the following sections.

\section{The stationary fields around a single Janus particle}
\label{sec:one-body_fields}

\subsection{The molecular concentrations}

The first step of the problem is to obtain the concentration profiles, because they are decoupled from the velocity field as a consequence of the fact that Eqs.~(\ref{Laplace-eq-ck}) for the concentrations and the boundary conditions (\ref{bc-ck}) and (\ref{bc-ck-inf}) do~not~depend on the velocity field.  The calculation is presented in App.~\ref{app:concentrations} by assuming that the Janus particle is oriented in the direction ${\bf u}={\bf 1}_z$.  The quantity entering the boundary conditions~(\ref{bc-v-2}) on the tangential components of the velocity field is thus given by
\be
\left( \sum_\varkappa b_\varkappa \boldsymbol{\nabla} c_\varkappa\right)_R = \frac{3}{2} \left( \boldsymbol{\mathsf 1} - {\bf n}{\bf n}\right) \cdot \left( b_{\rm A} {\bf g}_{\rm A} + b_{\rm B} {\bf g}_{\rm B}\right) + R \left(\frac{b_{\rm B}}{D_{\rm B}} - \frac{b_{\rm A}}{D_{\rm A}}\right) (\boldsymbol{\nabla} Z)_R \, ,
\label{sum-bk-gk}
\ee
where the function $Z$ obeys the equation $\nabla^2Z=0$ with the boundary conditions
\be
R(\partial_r Z)_R = \left[ {\rm Da} \, (Z)_R - \varrho - \frac{3}{2} \, R \, {\bf g}_\varrho \cdot {\bf n} \right] H^{\rm c}(\theta)
\label{Z}
\ee
and $(\boldsymbol{\nabla} Z)_\infty=0$, as expressed in terms of the dimensionless Damk\"ohler number
\be
{\rm Da} \equiv R \left(\frac{\kappa_+^{\rm c}}{D_{\rm A}} + \frac{\kappa_-^{\rm c}}{D_{\rm B}}\right) ,
\label{Da}
\ee
the reaction rate
\be
\varrho \equiv \kappa_+^{\rm c} \, c_{\rm A}^{\circ} - \kappa_-^{\rm c} \, c_{\rm B}^{\circ} \, ,
\label{varrho}
\ee
and its gradient
\be
{\bf g}_\varrho \equiv \kappa_+^{\rm c} \, {\bf g}_{\rm A} - \kappa_-^{\rm c} \, {\bf g}_{\rm B} \, .
\label{g_varrho}
\ee
The analytical expression of the function $Z$ is given in App.~\ref{app:concentrations}.

In the right-hand side of Eq.~(\ref{sum-bk-gk}), the first contribution comes from simple diffusiophoresis in the gradients of concentrations and the second is due to self-diffusiophoresis generated by the reaction.  In the absence of reaction (i.e., if $\kappa_\pm^{\rm c}=0$) or if the reaction is at equilibrium (i.e., if $\kappa_+^{\rm c} \, c_{\rm A}=\kappa_-^{\rm c} \, c_{\rm B}$), the second contribution is equal to zero.

The system evolves in the reaction-limited regime if ${\rm Da}\ll 1$, i.e., if $R\kappa_+^{\rm c} \ll D_{\rm A}$ and $R\kappa_-^{\rm c} \ll D_{\rm B}$; or in the diffusion-controlled regime if ${\rm Da}\gg 1$, i.e., if $R\kappa_+^{\rm c} \gg D_{\rm A}$ or $R\kappa_-^{\rm c} \gg D_{\rm B}$.

\subsection{The pressure and velocity fields}

Taking the divergence of Stokes' equations~(\ref{Stokes-eqs}) and using the incompressibility condition~(\ref{div.v=0}), we obtain the Laplace equation $\nabla^2p=0$ for the pressure.  Around a single Janus particle located at ${\bf r}=0$, the pressure can be expanded into spherical harmonics as
\be
p(r,\theta,\varphi) = \sum_{lm} p_{lm}(r) \, Y_{lm}(\theta,\varphi) \, ,
\label{pressure-SH}
\ee
where $l\ge \vert m \vert \ge 0$ and
\be
\frac{d^2}{dr^2}\left( r \, p_{lm}\right) - \frac{l(l+1)}{r}\, p_{lm} = 0 \, .
\ee
This linear second-order homogeneous ordinary differential equation is solved by considering the following linear superposition
\be
p_{lm}(r) = \eta_0 \left( a_{lm}\, r^l + b_{lm} \, r^{-l-1} \right) ,
\label{p_lm}
\ee
with some coefficients $a_{lm}$ and $b_{lm}$ to be determined.

The vector field of the fluid velocity around the Janus particle can be expanded as
\be
{\bf v}(r,\theta,\varphi) = \sum_{lm} \left[ f_{lm}(r) \, \bUps_{lm}(\theta,\varphi) + g_{lm}(r) \, \bPsi_{lm}(\theta,\varphi) + h_{lm}(r) \, \bPhi_{lm}(\theta,\varphi) \right]
\label{velocity-VSH}
\ee
into the vectorial spherical harmonics that are defined in App.~\ref{app:VSH} \cite{BEG85}.

In App.~\ref{app:velocity}, Stokes' equations~(\ref{Stokes-eqs}) and the incompressibility condition~(\ref{div.v=0}) are solved with the boundary conditions~(\ref{bc-v-1}), (\ref{bc-v-2}), and (\ref{bc-v-inf}) to obtain the radial functions $f_{lm}(r)$, $g_{lm}(r)$, and $h_{lm}(r)$ of the velocity field around a single Janus particle.  These radial functions are expressed as powers of the radial distance $r$ with coefficients given in terms of the velocity $v_i^{\circ}$ and velocity gradients $G_{ij}$, the particle velocity $V_i$ and angular velocity $\Omega_i$, and the molecular concentrations $c_\varkappa^{\circ}$ and concentration gradients ${\bf g}_\varkappa$ involved because of diffusiophoresis and the chemical reaction.  The analytical expressions of the radial functions are calculated in App.~\ref{app:velocity}.

Finally, we find that the disturbance velocity and pressure can be expressed as
\be
\Delta v_i = A_{jk} \left(\frac{r_j \, r_k}{r^5} - \frac{\delta_{jk}}{3 \, r^3} \right) r_i + O\left(\frac{1}{r^3}\right)
\qquad\mbox{and}\qquad
\Delta p = 2\eta_0 \, A_{jk} \left(\frac{r_j \, r_k}{r^5} - \frac{\delta_{jk}}{3 \, r^3} \right) + O\left(\frac{1}{r^4}\right) ,
\label{Dv-Dp-stresslet}
\ee
where the matrix of this stresslet has three parts
\be
A_{jk} = A_{jk}^{({\rm v})} + A_{jk}^{({\rm d})} + A_{jk}^{({\rm r})} \, ,
\label{A_jk}
\ee
given by Eqs.~(\ref{Dv-Dp-b_2m^v}), (\ref{A^d}), and~(\ref{A^r}), respectively.  The first part is proportional to the traceless tensor of velocity gradients, the second one is due to simple diffusiophoresis, and the third one to self-diffusiophoresis generated by the reaction.  These three parts contribute to the stress tensor of the active suspension according to Eq.~(\ref{Pij-suspension-LL}), as shown in the following section.

\section{The stress tensor of the dilute active suspension}
\label{sec:stress_tensor}

\subsection{The general expression}

According to Eqs.~(\ref{Pij-suspension-LL}) and~(\ref{coarse-graining}), the stress tensor of the dilute active suspension can be expressed as
\be
\overline{\sigma}_{ij} = -\overline{p}\, \delta_{ij} + \eta_0\, \overline{\nabla_i v_j+\nabla_j v_i} + \Delta \sigma_{ij}
\label{Pij-suspension}
\ee
with
\be
\Delta \sigma_{ij} = \int d^2u\, f_{\rm C} \int \left\{- \Delta p  \, r_j \, d\Sigma_i + \eta_0 \left[r_j \left( \nabla_i \Delta v_k + \nabla_k \Delta v_i\right) d\Sigma_k - \Delta v_i \, d\Sigma_j - \Delta v_j \, d\Sigma_i \right]\right\} 
\label{DPij-suspension-LL}
\ee
in terms of the disturbance velocity and pressure given by Eq.~(\ref{Dv-Dp-stresslet}).  The calculation of Eq.~(\ref{DPij-suspension-LL}) for the part of the stress tensor due to the colloidal particles is carried out in App.~\ref{app:P_ij}.  This calculation shows that
\be
\Delta \sigma_{ij} = -\frac{4\pi}{3} \, \eta_0 \int d^2u \, f_{\rm C} \left( A_{ij}+A_{ji} - \frac{2}{3} \, A_{kk} \, \delta_{ij}\right)
\label{P_ij-A_ij}
\ee
in terms of the stresslet matrix~(\ref{A_jk}).  The three parts of this matrix lead to three 
corresponding contributions to the stress tensor,
\be
\Delta \sigma_{ij} = \Delta \sigma_{ij}^{({\rm v})} + \Delta \sigma_{ij}^{({\rm d})} + \Delta \sigma_{ij}^{({\rm r})} \, ,
\label{DP_ij-vdr}
\ee
which are given here below, using the results of the calculations carried out in App.~\ref{app:velocity}.

\subsection{The contribution to the shear viscosity}

Inserting the stresslet matrix given in Eq.~(\ref{Dv-Dp-b_2m^v}) into Eq.~(\ref{P_ij-A_ij}) and using the property that ${\rm tr}\, \boldsymbol{\mathsf A}^{({\rm v})}=A_{kk}^{({\rm v})}=0$ show that
\be
\Delta \sigma_{ij}^{({\rm v})} = \frac{5}{2} \, \frac{\displaystyle 1 + \frac{2b}{R}}{\displaystyle 1 + \frac{5b}{R}} \, \phi \, \eta_0 \left( G_{ij}+G_{ji}\right) , 
\qquad\mbox{where}\qquad
\phi\equiv \frac{4\pi}{3}\, R^3 \, n_{\rm C}
\label{DP_ij-v}
\ee
is the volume fraction occupied by the Janus particles in the suspension. Since $G_{ij}=\overline{\nabla_j v_i}$, this contribution to the stress tensor~(\ref{Pij-suspension}) implies that the effective shear viscosity of the suspension is given by
\be
\eta = \eta_0 \left(1 + \frac{5}{2} \, \frac{\displaystyle 1 + \frac{2b}{R}}{\displaystyle 1 + \frac{5b}{R}} \, \phi \right) ,
\label{eta}
\ee
at leading order in the colloidal fraction, which is a well-known result since Einstein's pioneering work \cite{E1906,LP08,PW22}.

\subsection{The contribution from simple diffusiophoresis}

Here, we consider the contribution from simple diffusiophoresis given by the stresslet matrix~(\ref{A^d}).  Inserting it into Eq.~(\ref{P_ij-A_ij}) leads to
\be
\Delta \sigma_{ij}^{({\rm d})} = \frac{15\pi}{8} \, \frac{R^2  \eta_0}{\displaystyle 1 + \frac{5b}{R}} \int d^2u \, f_{\rm C} \sum_\varkappa \Delta b_\varkappa \left[ g_{\varkappa i} \, u_j + g_{\varkappa j} \, u_i+ {\bf g}_\varkappa \cdot{\bf u} \left( u_i \, u_j - \delta_{ij} \right)\right] ,
\label{DP_ij-d}
\ee
where ${\bf g}_\varkappa=\overline{\boldsymbol{\nabla} c_\varkappa}$ are the molecular concentration gradients defined on the scale of the coarse-grained description and $\Delta b_\varkappa= b_\varkappa^{\rm c} - b_\varkappa^{\rm n}$ are the differences of diffusiophoresis constants between the catalytic and non-catalytic hemispheres of the Janus particles.  Therefore, this contribution is not equal to zero if there is heterogeneity in the diffusiophoretic effect at the surface of the particles.  We note that this contribution to the stress tensor is symmetric, $\Delta\sigma_{ij}^{({\rm d})}=\Delta\sigma_{ji}^{({\rm d})}$, and traceless, ${\rm tr}\, \Delta\boldsymbol{\sigma}^{({\rm d})}=\Delta\sigma_{ii}^{({\rm d})}=0$, as expected for torque-free rigid Janus particles.

Using the decomposition~(\ref{f_C-decomp}) of the distribution function of the Janus particles into the moments of their orientation (\ref{p_i})-(\ref{r_ijk}), we find that
\be
\Delta \sigma_{ij}^{({\rm d})} = \frac{9\pi}{4} \, \frac{R^2 \eta_0}{\displaystyle 1 + \frac{5b}{R}} \sum_\varkappa \Delta b_\varkappa \left( g_{\varkappa i} \, p_j + g_{\varkappa j} \, p_i - \frac{2}{3} \, {\bf g}_\varkappa \cdot {\bf p} \, \delta_{ij} + \frac{5}{6} \, g_{\varkappa k} \, r_{ijk} \right) ,
\label{DP_ij-d-r_ijk}
\ee
as expressed in terms of the concentration gradients $g_{\varkappa k}=\overline{\nabla_k c_\varkappa}$ of the molecular species $\varkappa\in\{{\rm A},{\rm B}\}$, the vector field~(\ref{p_i}), and the rank-three tensor~(\ref{r_ijk}) of the particle orientation.

\subsection{The contribution from reaction}

Finally, the contribution from the reaction can be obtained by inserting the stresslet matrix~(\ref{A^r}) into the formula~(\ref{P_ij-A_ij}), giving
\be
\Delta \sigma_{ij}^{({\rm r})} = \frac{4\pi R^2 \eta_0}{\displaystyle 1 + \frac{5b}{R}} \int d^2u \, f_{\rm C} \left[\left(\alpha_\Lambda \, \varrho + \frac{3R}{2} \, \beta_\Lambda \, {\bf g}_{\varrho}\cdot{\bf u}\right) \left( 3 \, u_i \, u_j -\delta_{ij} \right) + \frac{3R}{2} \, \gamma_\Lambda \left( g_{\varrho i} \, u_j+ g_{\varrho j}\, u_i - 2 \, {\bf g}_\varrho \cdot{\bf u} \, u_i \, u_j \right)\right]
\label{DP_ij-r}
\ee
in terms of the reaction rate~(\ref{varrho}) with the average molecular concentrations $c_\varkappa^{\circ}=\overline{c_\varkappa}$, its gradient~(\ref{g_varrho}) with ${\bf g}_\varkappa=\overline{\boldsymbol{\nabla} c_\varkappa}$, and the coefficients~(\ref{alpha_Lambda})-(\ref{gamma_Lambda}).  This contribution due to the self-diffusiophoresis generated by the reaction is also symmetric $\Delta\sigma_{ij}^{({\rm r})}=\Delta\sigma_{ji}^{({\rm r})}$, and traceless, ${\rm tr}\, \Delta\boldsymbol{\sigma}^{({\rm r})}=\Delta\sigma_{ii}^{({\rm r})}=0$.

Using the decomposition~(\ref{f_C-decomp}), this contribution can be expressed as follows
\be
\Delta \sigma_{ij}^{({\rm r})} = \frac{12 \pi R^2 \eta_0}{\displaystyle 1 + \frac{5b}{R}} \left[ \alpha_\Lambda \, \varrho \, q_{ij} + \frac{3}{10} \left(\beta_\Lambda + \gamma_\Lambda\right) R \left( g_{\varrho i} \, p_j + g_{\varrho j} \, p_i - \frac{2}{3} \, {\bf g}_{\varrho} \cdot {\bf p} \, \delta_{ij} \right) + \left(\frac{3}{2} \, \beta_\Lambda -\gamma_\Lambda\right) R \, g_{\varrho k} \, r_{ijk} \right]
\label{DP_ij-r-p_i-q_ij-r_ijk}
\ee
in terms of the orientation tensors of rank~one~(\ref{p_i}), rank~two~(\ref{q_ij}), and rank~three~(\ref{r_ijk}).  We note that the existence of the first term going as $\varrho\, q_{ij}$ has been anticipated in Ref.~\cite{R10}.

\section{The case of uniform diffusiophoretic constants}
\label{sec:uniform-case}

In the case where the diffusiophoretic constants~(\ref{diffusio-coeff}) take the same values on both hemispheres, i.e., $\Delta b_\varkappa =0$ for $\varkappa\in\{{\rm A},{\rm B}\}$, the contribution~(\ref{DP_ij-d-r_ijk}) from simple diffusiophoresis is equal to zero.

In the contribution~(\ref{DP_ij-r-p_i-q_ij-r_ijk}) from reaction, the coefficients~(\ref{alpha_Lambda}), (\ref{beta_Lambda}), and~(\ref{gamma_Lambda}) reduce to
\be
\alpha_\Lambda = \Lambda \, a_2 \, , \qquad
\beta_\Lambda = \Lambda \, b_2 \, , \qquad
\gamma_\Lambda = \Lambda \, c_2 \, , \qquad
\mbox{with}\qquad
\Lambda \equiv \frac{b_{\rm B}}{D_{\rm B}} - \frac{b_{\rm A}}{D_{\rm A}} 
\label{uniform-Lambda}
\ee
according to Eq.~(\ref{Lambda^h}).  Furthermore, the reaction rate~(\ref{varrho}) and its gradient~(\ref{g_varrho}) can be expressed in terms of the particle velocity generated by the reaction, which is known to take the following value~\cite{GK20},\footnote{In Ref.~\cite{GK20}, this velocity was denoted $V_{\rm sd}\equiv V^{({\rm r})}$ to refer to its self-diffusiophoretic origin.}
\be
V^{({\rm r})} = \frac{2\, a_1}{\displaystyle 3\left(1+\frac{2b}{R}\right)} \, \Lambda \, \varrho \, ,
\label{V^r}
\ee
as summarised in App.~\ref{app:velocities-uniform}.
If, moreover, we introduce the local mean values
\be
\langle u_i \rangle \equiv \frac{p_i}{n_{\rm C}} \, , \qquad
\langle Q_{ij} \rangle \equiv \frac{q_{ij}}{n_{\rm C}} \, , \qquad\mbox{and}\qquad
\langle R_{ijk} \rangle \equiv \frac{r_{ijk}}{n_{\rm C}} 
\ee
in terms of the moments~(\ref{p_i})-(\ref{r_ijk}), the contribution~(\ref{DP_ij-r-p_i-q_ij-r_ijk}) from reaction to the stress tensor can be equivalently written as
\bea
&& \Delta \sigma_{ij}^{({\rm r})} = \frac{27}{2 a_1} \, \frac{\displaystyle 1 + \frac{2b}{R}}{\displaystyle 1 + \frac{5b}{R}} \, \phi \, \eta_0 \bigg[ a_2 \, \frac{V^{({\rm r})}}{R}  \, \langle Q_{ij}\rangle + \frac{3}{10} \, (b_2 + c_2) \left( \nabla_i V^{({\rm r})}  \, \langle u_j\rangle + \nabla_j V^{({\rm r})}  \, \langle u_i\rangle  - \frac{2}{3} \, \boldsymbol{\nabla} V^{({\rm r})} \cdot \langle{\bf u}\rangle \, \delta_{ij} \right) \nonumber\\
&&\qquad\qquad\qquad\qquad\qquad\qquad\qquad\qquad\qquad
 + \left(\frac{3}{2} \, b_2 - c_2\right) \nabla_k V^{({\rm r})} \, \langle R_{ijk}\rangle \bigg] \, ,
\label{DP_ij-r-u_i-Q_ij-R_ijk}
\eea
where $\phi$ is the volume fraction of the Janus particles in the suspension, as introduced in Eq.~(\ref{DP_ij-v}).

\begin{figure}[h]
\centerline{\scalebox{0.65}{\includegraphics{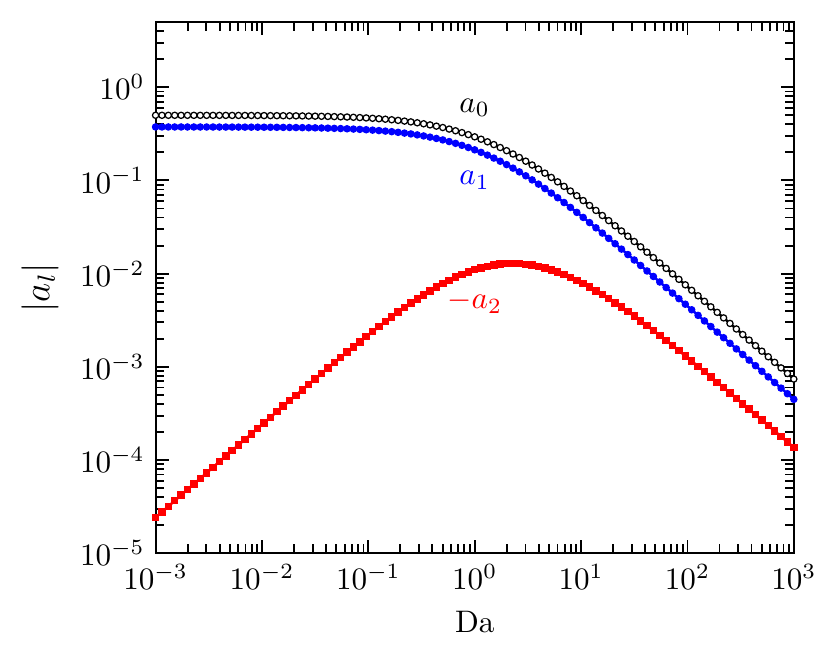}}}
\caption{Plots of the values of the coefficients $a_0$, $a_1$, and $a_2$ versus the Damk\"ohler number~(\ref{Da}).}
\label{fig2}
\end{figure}

\begin{figure}[h]
\centerline{\scalebox{0.65}{\includegraphics{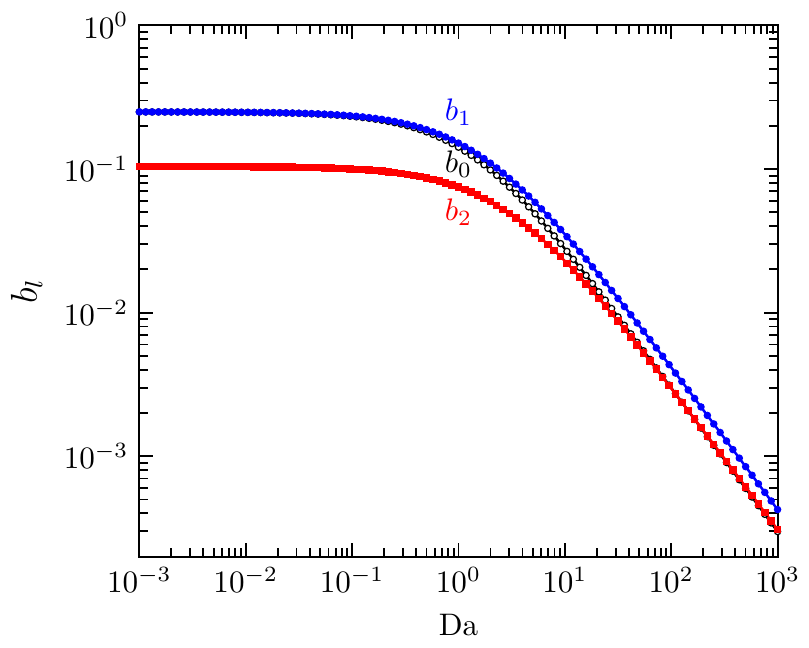}}}
\caption{Plots of the values of the coefficients $b_0$, $b_1$, and $b_2$ versus the Damk\"ohler number~(\ref{Da}).}
\label{fig3}
\end{figure}

\begin{figure}[h]
\centerline{\scalebox{0.65}{\includegraphics{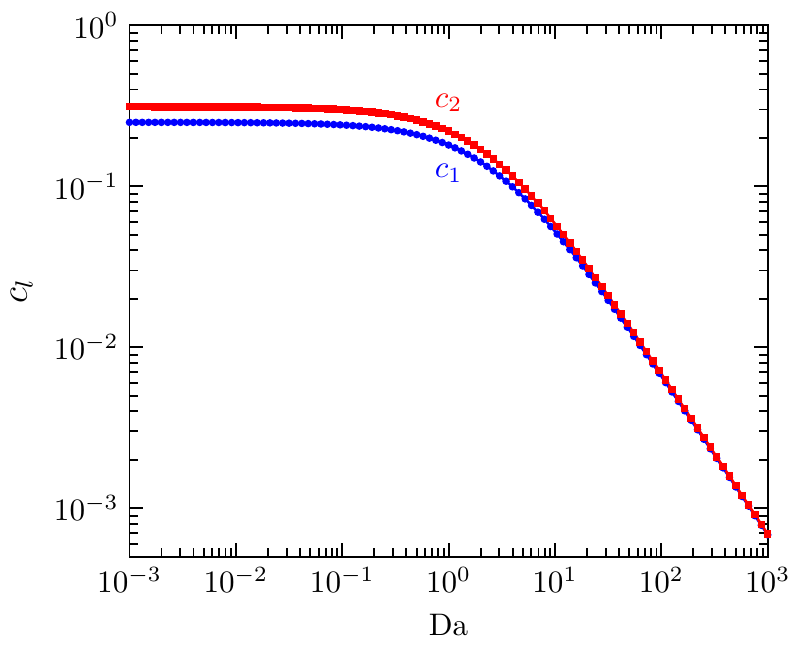}}}
\caption{Plots of the values of the coefficients $c_1$ and $c_2$ versus the Damk\"ohler number~(\ref{Da}).}
\label{fig4}
\end{figure}

We note that, under the same conditions, the overall reaction rate of a single Janus particle is given by Eq.~(\ref{W-single-particle}) in terms of the coefficients $a_0$ and $b_0$; and its vectorial velocity by Eq.~(\ref{V-single-particle}) in terms of the coefficients $a_1$, $b_1$, and $c_1$ \cite{GK20}.  The coefficients $a_0$, $a_1$, $a_2$, $b_0$, $b_1$, $b_2$, $c_1$, and $c_2$ can be calculated with Eqs.~(\ref{a_l-b_l-c_l})-(\ref{M-N}) using {\tt Mathematica}~\cite{W88}.  Their values are given in Table~\ref{TableI} and plotted in Figs.~\ref{fig2}-\ref{fig4}, as a function of the Damk\"ohler number~(\ref{Da}).  In the diffusion-controlled regime, they decrease as ${\rm Da}^{-1}$ for ${\rm Da} \gg 1$.  We note that the coefficient~$a_2$ is negative and equal to zero at ${\rm Da}=0$.  Moreover, we see in Fig.~\ref{fig2} that the coefficient~$a_2$ goes to zero for ${\rm Da}\to 0$ in a way that is proportional to ${\rm Da}$.

\begin{table}[h]
\caption{\label{table} Values of the coefficients $a_0$, $a_1$, $a_2$, $b_0$, $b_1$, $b_2$, $c_1$, and $c_2$ versus the Damk\"ohler number~(\ref{Da}) computed with Eqs.~(\ref{a_l-b_l-c_l})-(\ref{M-N}) truncated to $l_{\rm max}=100$.}
\label{TableI}
\vspace{5mm}
\begin{center}
\begin{tabular}{|ccccccccc|}
\hline
{\rm Da} & $a_0$ & $a_1$ & $a_2$ & $b_0$ & $b_1$& $b_2$ & $c_1$ & $c_2$ \\
\hline
0   & 0.5             &    0.375       &        0.0           &   0.25         &   0.25         &  0.10417   &   0.25         &   0.3125    \\
1   &	0.29348     &    0.21286   &   $-$0.01102   &   0.14190   &   0.15160   &  0.07485   &   0.18057   &   0.22005  \\
2   &   0.20824   &    0.14763   &    $-$0.01279	&   0.09842   &   0.10974   &  0.05920   &  0.14158   &   0.16932  \\
3   &   0.16157   &    0.11265   &    $-$0.01261  &   0.07510   &   0.08632   &  0.04917   &   0.11656   &   0.13737  \\
4   &   0.13208   &    0.09092   &   $-$0.01193   &   0.06061   &   0.07127   &  0.04214   &   0.09911   &   0.11545  \\
5   &   0.11175    &    0.07613   &   $-$0.01115   &   0.05075   &   0.06075   &  0.03690   &   0.08625   &   0.09950  \\
6   &   0.09686   &    0.06544   &   $-$0.01040   &   0.04362   &   0.05297   &  0.03284   &   0.07636   &   0.08737  \\
7   &   0.08549   &    0.05735   &   $-$0.00971   &   0.03823   &   0.04698   &  0.02960   &   0.06852   &   0.07786  \\
8   &   0.07653   &    0.05102   &   $-$0.00909   &   0.03402   &   0.04222   &  0.02695   &   0.06215   &   0.07020  \\
9   &   0.06927   &    0.04594   &   $-$0.00853   &   0.03063   &   0.03834   &  0.02474   &   0.05688   &   0.06389  \\
10 &   0.06327   &    0.04177   &   $-$0.00803   &   0.02785   &   0.03512   &  0.02287   &   0.05243   &   0.05862  \\
\hline
\end{tabular}
\end{center}
\end{table}

The contribution~(\ref{DP_ij-r-u_i-Q_ij-R_ijk}) from the reaction can be compared with the contribution~(\ref{DP_ij-v}) to shear viscosity.  They would have comparable values if the gradients $G_{ij}$ of the fluid velocity would be of the same order of magnitude as the particle velocity~(\ref{V^r}) divided by the particle radius $R$ or its gradient components $\nabla_i V^{({\rm r})}$, and if the order parameters $\langle Q_{ij}\rangle$ or $\langle u_i\rangle$ of the particle orientation were maximal.  Generally speaking, Eq.~(\ref{DP_ij-r-u_i-Q_ij-R_ijk}) is the analogue of the force-velocity relation characterising active systems such as muscles.

To evaluate the contribution~(\ref{DP_ij-r-u_i-Q_ij-R_ijk}), we may assume that the Navier slip length is much smaller than the particle radius (i.e., $b=0$)\footnote{The condition $b\ll R$ is indeed satisfied for common solid surfaces in contact with aqueous solutions \cite{JYB06,BTB08,COCMS21}.} and that the terms with the components of the gradient $\boldsymbol{\nabla}V^{({\rm r})}$ are negligible.  In this case, we have that 
\be
\Delta\sigma_{ij}^{({\rm r})} \simeq \frac{27 a_2}{2 a_1} \, \phi \, \eta_0 \, \frac{V^{({\rm r})}}{R}  \, \langle Q_{ij}\rangle = 9 \, a_2 \, \phi \, \eta_0 \, R^{-1} \, \Lambda \varrho \, \left( \langle u_i u_j \rangle - \frac{1}{3}\, \delta_{ij}\right) ,
\ee
because the particle velocity~(\ref{V^r}) is given by $V^{({\rm r})}\simeq (2/3) a_1 \Lambda \varrho$ in the same approximation.  An active particle is said to be a pusher if $\Lambda\varrho >0$ and a puller if $\Lambda\varrho <0$ \cite{RHSK17,GK19}.\footnote{The notation $\Upsilon\equiv\Lambda\varrho$ is used in Ref.~\cite{GK19}.}

Now, if the particles were fully polarized in the $x$-direction with ${\bf u}=(1,0,0)$, the activity would change the pressure by $\Delta P_{xx}^{({\rm r})}=-\Delta\sigma_{xx}^{({\rm r})}\simeq -6 a_2 \phi\eta_0 R^{-1}\Lambda\varrho$ in that direction and by $\Delta P_{yy}^{({\rm r})}=\Delta P_{zz}^{({\rm r})}=-\frac{1}{2}\, \Delta P_{xx}^{({\rm r})}$ in the orthogonal directions to maintain the incompressibility.  Since $a_2 <0$, the medium is extensile with $\Delta P_{xx}^{({\rm r})}>0$ if the active particles are pushers and the medium is contractile with $\Delta P_{xx}^{({\rm r})}<0$ if the active particles are pullers, which is consistent with expectations \cite{R10}.  In an active suspension of Damk\"ohler number equal to ${\rm Da}=10$ composed of particles of radius $R=1\,\mu$m, moving with the speed $V^{({\rm r})}=10\,\mu$m/s in an aqueous solution of viscosity $\eta_0=10^{-3}\,$Pa~s, and occupying a volume fraction $\phi=0.1$, the activity would generate a change of pressure of about $\Delta P_{xx}^{({\rm r})} \simeq 1.7\times 10^{-3}$~Pa.  This effect is proportional to the particle velocity $V^{({\rm r})}$ and to their volume fraction $\phi$.

\section{Conclusion and perspectives}
\label{sec:conclusion}

In this paper, the method pioneered by Einstein \cite{E1906,LL59,B70} to calculate the stress tensor was extended from passive to active dilute suspensions composed of Janus colloidal particles, which are propelled by self-diffusiophoresis and powered with chemical energy.  In the dilute-system limit, the particles may be considered to be far apart and isolated from each other, so that their mutual interactions can be neglected.  Under such circumstances, the stress tensor of the suspension can be deduced from the velocity and pressure of the fluid around a single Janus particle moving in coarse-grained background fields.

The particles were supposed to be spherical and made of two hemispheres: a catalytic hemisphere, where the chemical reaction powering the activity occurs, and the other one being non-catalytic.  The propulsion of the particles was assumed to be induced by diffusiophoresis locally coupling the fluid velocity with the molecular concentration gradients generated by the surface chemical reaction on the catalytic hemisphere of the Janus particles.  Accordingly, the molecular concentrations were first obtained before calculating the one-particle velocity field.  The latter was expanded into vectorial spherical harmonics and corresponding radial functions, which were calculated by solving the steady Stokes equations and the condition of incompressibility.  The coefficients of the radial functions were determined using the boundary conditions at the surface of the Janus particle and far away from the particle, together with the force-free and torque-free conditions.  

As a consequence, the velocity and pressure disturbance fields of a single particle behave as $\Delta{\bf v}=O(r^{-2})$ and $\Delta p=O(r^{-3})$, leading to three contributions to the stress tensor due to the presence of Janus particles in the suspension.  In general, these contributions depend on the orientations of the Janus particles and their distribution.  All these contributions to the stress tensor are traceless and symmetric in consistency with the incompressibility and torque-free conditions, respectively.

The first contribution leads to an enhancement of the shear viscosity, which is proportional to the volume fraction of suspended particles.  This contribution is well known since Einstein's classic work \cite{E1906}.  Since the velocity field was here assumed to satisfy partial slip boundary conditions, the effective shear viscosity was here obtained with its dependence on the Navier slip length, generalizing Einstein's result for no-slip spheres \cite{LP08,PW22}.

The second contribution is due to the simple diffusiophoresis of the particles in the gradients of molecular concentrations, which would persist in the absence of chemical reactions, i.e., if the suspension was passive.  However, this contribution only exists if the diffusiophoretic constants take different values on the materials composing the two hemispheres of the Janus particles.

The third contribution constitutes the active part of the stress tensor, because it is proportional to the reaction rate and its gradient, which are equal to zero at chemical thermodynamic equilibrium.  This contribution from the activity of the suspension also depends on the orientation of the Janus particles.  Since the stress tensor is a rank-two tensor and the reaction rate density is a scalar, the corresponding term is proportional to the tensor characterising the nematic order of the Janus particles.  Terms coupling the gradient of the reaction rate with the vector of polar order are also identified.

The three contributions to the stress tensor are proportional to the volume fraction occupied by the suspended Janus particles. Therefore, their effects are small in dilute suspensions.  Nevertheless, the systematic calculation achieved in this paper predicts the magnitude of these effects and these quantitative predictions can be tested experimentally.   Furthermore, these effects are expected to become larger in denser suspensions and, in particular, in strongly heterogeneous suspensions, having undergone some clustering instability.  In this regard, the contributions of higher powers in the volume fraction could be obtained by extending the systematic methods developed in Ref.~\cite{BKM77} from passive to active suspensions.

Similar calculations can be carried out for active particles with other geometries like spheroidal particles with a varying size of the catalytic patch \cite{PDTR10} or sphere dimers with a catalytic sphere bonded to a non-catalytic one \cite{RK15}.

In a more general perspective, the expression that was here derived for the stress tensor establishes the hydrodynamic equations for the coarse-grained velocity field in the suspension.  These equations should be supplemented with other equations for the coarse-grained molecular concentrations and the mean distribution function of the Janus particles $f_{\rm C}({\bf r},{\bf u},t)$, in order to obtain the complete set of equations ruling the time evolution of the suspension at the macroscale.

The results provide a fundamental justification to the analytic expressions commonly assumed for the stress tensor in the theories describing suspensions as continuous media.  They also show that the study of synthetic active matter can open interesting perspectives to understand analogue phenomena in biological systems.  In particular, the sarcomers composing muscles have a head-to-head structure with nematic ordering in the muscular fibers.  In this analogy, the force exerted by muscles can be understood in terms of an active stress tensor similar to the one that is here obtained on the basis of the physico-chemical principles.  Along these lines, the present study may also lead to a better understanding of the active stress tensor for the mathematical description of biological tissues.

The present calculations leave open the issue of external forces or external torques exerted on the suspended particles.  The presence of external forces should add a corresponding force density in the coarse-grained Navier-Stokes equations, while external torques should contribute by an antisymmetric part to the effective stress tensor \cite{LL59,B70}.

Finally, we note that the mesoscopic approach developed in this paper is based on the knowledge of the constitutive properties of the solution and the fluid-solid interfaces.  In turn, these properties should be determined by the microscopic dynamics of the atoms and molecules composing the system.  Therefore, the fully microscopic derivation of the stress tensor in active suspensions will require first-principles approaches based on non-equilibrium statistical mechanics \cite{RSGK20,RSK24,G22}.

\section*{Acknowledgments}

This research was supported by the Universit\'e Libre de Bruxelles (ULB).
The Author thanks Raymond Kapral for discussions and David Gaspard for expertise support during this work.

\vskip 0.5 cm

\appendix

\section{The stationary one-particle profiles of molecular concentrations}
\label{app:concentrations}

In this appendix, the equations~(\ref{Laplace-eq-ck}) with the boundary conditions~(\ref{bc-ck}) and~(\ref{bc-ck-inf}) are solved to obtain the stationary molecular concentrations around one Janus particle located at ${\bf r}=0$ and oriented vertically with ${\bf u}={\bf 1}_z$.  The geometry of the particle and the spherical coordinates are shown in Fig.~\ref{fig3}.  The methods of Ref.~\cite{GK20} are used in the following calculation.

\begin{figure}[h]
\centerline{\scalebox{1.0}{\includegraphics{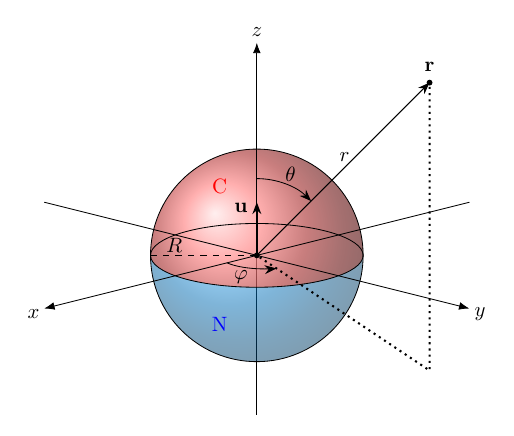}}}
\caption{Schematic representation of one Janus particle and the used spherical coordinates.}
\label{fig5}
\end{figure}

Since ${\bf n}\cdot\boldsymbol{\nabla}=\partial_r$ because ${\bf n}={\bf 1}_r={\bf r}/r$ is the unit vector in the radial direction, the concentrations of molecular species ${\rm A}$ and ${\rm B}$ satisfy the following equations and boundary conditions,
\bea
&&\nabla^2 c_{\rm A}=0 \, , \qquad D_{\rm A} \left(\partial_r c_{\rm A}\right)_R = + \left( \kappa_+^{\rm c} c_{\rm A} - \kappa_-^{\rm c} c_{\rm B}\right)_R H^{\rm c}(\theta) \, , \qquad \left(\boldsymbol{\nabla} c_{\rm A}\right)_\infty = {\bf g}_{\rm A} \, , \\
&&\nabla^2 c_{\rm B}=0 \, , \qquad D_{\rm B} \left(\partial_r c_{\rm B}\right)_R = - \left( \kappa_+^{\rm c} c_{\rm A} - \kappa_-^{\rm c} c_{\rm B}\right)_R H^{\rm c}(\theta) \, , \qquad \left(\boldsymbol{\nabla} c_{\rm B}\right)_\infty = {\bf g}_{\rm B} \, ,
\eea
where $H^{\rm c}(\theta)=1$ on the catalytic hemisphere and $H^{\rm c}(\theta)=0$ on the non-catalytic one.
The molecular concentrations can be decomposed as
\be\label{cA-cB-phi-psi}
c_{\rm A} = \frac{1}{D_{\rm A}} \left( \frac{\kappa_-^{\rm c}}{D_{\rm B}} \, \phi + \frac{1}{\ell}\, \psi\right) 
\qquad\mbox{and}\qquad
c_{\rm B} = \frac{1}{D_{\rm B}} \left( \frac{\kappa_+^{\rm c}}{D_{\rm A}} \, \phi - \frac{1}{\ell}\, \psi\right) \qquad\mbox{with}\qquad 
\ell \equiv \left(\frac{\kappa_+^{\rm c}}{D_{\rm A}} + \frac{\kappa_-^{\rm c}}{D_{\rm B}}\right)^{-1} = \frac{R}{\rm Da} 
\ee
into the following functions
\be\label{phi-psi}
\phi \equiv \ell (D_{\rm A}\, c_{\rm A} + D_{\rm B}\, c_{\rm B})
\qquad\mbox{and}\qquad
\psi \equiv \ell^2 (\kappa_+^{\rm c}\, c_{\rm A} -\kappa_-^{\rm c}\, c_{\rm B}) \, , 
\ee
obeying
\bea
&&\nabla^2 \phi=0 \, , \qquad \left(\partial_r \phi\right)_R = 0 \, , \qquad\qquad\qquad\qquad\quad \left(\boldsymbol{\nabla} \phi \right)_\infty = {\bf g}_{\phi} \, , \\
&&\nabla^2 \psi=0 \, , \qquad R\left(\partial_r \psi\right)_R = {\rm Da} \left(\psi\right)_R H^{\rm c}(\theta) \, , \qquad \left(\boldsymbol{\nabla} \psi \right)_\infty = {\bf g}_{\psi} \, , \label{eqs_for_psi}
\eea
where
\be
{\bf g}_{\phi} \equiv \ell (D_{\rm A}\, {\bf g}_{\rm A} + D_{\rm B}\, {\bf g}_{\rm B})
\qquad\mbox{and}\qquad
{\bf g}_{\psi} \equiv \ell^2 (\kappa_+^{\rm c}\, {\bf g}_{\rm A} -\kappa_-^{\rm c}\, {\bf g}_{\rm B}) = \ell^2 \, {\bf g}_\varrho \, .
\ee
The solutions of these two problems are given by
\be
\phi=\phi^{\circ} + {\bf g}_{\phi}\cdot{\bf r} \left( 1 + \frac{R^3}{2r^3}\right)
\label{phi}
\ee
and
\be
\psi=\ell^2 \left[ \varrho + {\bf g}_\varrho\cdot{\bf r} \left( 1 + \frac{R^3}{2r^3}\right) - {\rm Da} \, Z \right]
\label{psi}
\ee
in terms of the reaction rate~(\ref{varrho}), its gradient~(\ref{g_varrho}), and the function $Z(r,\theta,\varphi)$ that is the solution of the problem~(\ref{Z}) as a consequence of Eq.~(\ref{eqs_for_psi}).\footnote{The notations used in Ref.~\cite{GK20} are $(\Phi)_{2020}=\phi$, $(\Psi)_{2020}=\psi$, $(\Psi_0)_{2020}=\ell^2\varrho$, and $({\cal F})_{2020}=\varrho^{-1}Z$.} The latter can be expanded as
\be
Z(r,\theta,\varphi) = \sum_{lm} z_{lm} \left(\frac{R}{r}\right)^{l+1} Y_{lm}(\theta,\varphi)
\label{Z-expansion}
\ee
into the spherical harmonics defined by
\be
Y_{lm}(\theta,\varphi) =(-1)^m \sqrt{\frac{(2l+1)(l-m)!}{4\pi (l+m)!}} P_l^m(\cos\theta) \, {\rm e}^{\imath m \varphi} 
\qquad\mbox{and}\qquad Y_{l,-m}(\theta,\varphi) = (-1)^m Y_{lm}^*(\theta,\varphi) \qquad\mbox{for}\qquad m\ge 0
\ee
in terms of the associated Legendre functions $P_l^{\vert m \vert}(\xi)=(1-\xi^2)^{\vert m \vert/2} \frac{d^{\vert m \vert}}{d\xi^{\vert m \vert}} P_l(\xi)$, where $P_l(\xi)$ is the Legendre polynomial of degree $l \ge \vert m \vert \ge 0$ \cite{BJ00}.

The coefficients $z_{lm}$ can be calculated by inserting the expansion~(\ref{Z-expansion}) into the boundary conditions of the problem~(\ref{Z}) and by using the orthonormality of the spherical harmonics, $\int Y_{lm}^* Y_{l'm'} d\Omega=\delta_{ll'} \delta_{mm'}$, leading to
\bea
&& z_{l0} = \sqrt{\frac{4\pi}{2l+1}} \left( a_l \, \varrho + \frac{3R}{2} \, b_l \, g_{\varrho z} \right) \qquad\mbox{for}\qquad l\ge 0 \, , \label{z_l0}\\
&& z_{l1}=-z_{l,-1}^* = \sqrt{\frac{4\pi}{2l+1}} \frac{3R}{2} \, \frac{c_l}{\sqrt{l(l+1)}} \, (-g_{\varrho x}+\imath\, g_{\varrho y} ) \qquad\mbox{for}\qquad l\ge 1 \, , \qquad\mbox{and}  \label{z_l1}\\
&& z_{lm}=0 \qquad \mbox{for}\qquad  l \ge \vert m \vert \ge 2 \, ,
\label{z_lm}
\eea
where the coefficients are given by\footnote{The present quantities $c_l$, ${\cal C}_l$, and $N_{ll'}$ are related to those used in the 2020 paper~\cite{GK20} according to $c_l=(c_l)_{2020}\sqrt{l(l+1)/2}$, ${\cal C}_l=({\cal C}_l)_{2020}/\sqrt{2l(l+1)}$, and $N_{ll'}=(N_{ll'})_{2020}/\sqrt{l(l+1)l'(l'+1)}$.  The other quantities are the same.  This change in the definition of the coefficients $c_l$ allows us to have similar expansions for $\gamma_\Lambda$ in Eq.~(\ref{gamma_Lambda}) as for $\alpha_\Lambda$ and $\beta_\Lambda$ in Eqs.~(\ref{alpha_Lambda}) and~(\ref{beta_Lambda}).}
\be
a_l \equiv \left({\boldsymbol{\mathsf M}}^{-1}\cdot\pmb{\cal A}\right)_l \, , \qquad
b_l \equiv \left({\boldsymbol{\mathsf M}}^{-1}\cdot\pmb{\cal B}\right)_l \, , \qquad
c_l \equiv  \left({\boldsymbol{\mathsf N}}^{-1}\cdot\pmb{\cal C}\right)_l 
\label{a_l-b_l-c_l}
\ee
with
\be
{\cal A}_l \equiv \int _0^1 d\xi \, P_l(\xi) \, , \qquad {\cal B}_l \equiv \int _0^1 d\xi \, P_1(\xi) \, P_l(\xi) \, , \qquad
{\cal C}_l \equiv \int _0^1 d\xi \, \frac{P_1^1(\xi)}{2} \, \frac{P_l^1(\xi)}{l(l+1)} \, , 
\label{A_l-B_l-C_l}
\ee
\be
M_{ll'} \equiv 2 \, \frac{l+1}{2l+1} \, \delta_{ll'} + {\rm Da} \, \int_0^1 d\xi \, P_l(\xi)\, P_{l'}(\xi) \, , 
\qquad\mbox{and}\qquad
N_{ll'} \equiv \frac{2}{l(2l+1)} \, \delta_{ll'} + {\rm Da} \, \int_0^1 d\xi \, \frac{P_l^1(\xi)}{l(l+1)} \, \frac{P_{l'}^1(\xi)}{l'(l'+1)} \, . 
\label{M-N}
\ee

We note that the function $Z$ can also be decomposed as
\be
Z = \varrho \, K + \frac{3}{2} \, R \, {\bf g}_\varrho\cdot{\bf L} 
\label{Z-K-L}
\ee
into the functions
\bea
&& K \equiv \sum_{l=0}^{\infty} a_l \, P_l(\cos\theta) \left(\frac{R}{r}\right)^{l+1} , \label{K}\\
&& L_z \equiv \sum_{l=0}^{\infty} b_l \, P_l(\cos\theta) \left(\frac{R}{r}\right)^{l+1} , \label{Lz}\\
&& \frac{L_x}{\cos\varphi} = \frac{L_y}{\sin\varphi} \equiv 2 \sum_{l=1}^{\infty} c_l \, \frac{P_l^1(\cos\theta)}{l(l+1)} \left(\frac{R}{r}\right)^{l+1} . \label{Lx-Ly}
\eea

The knowledge of these functions allows us to calculate the overall reaction rate~(\ref{overall_reaction_rate}) for a single Janus particle. Indeed, the molecular concentrations can be expressed by Eq.~(\ref{cA-cB-phi-psi}) in terms of the functions defined in Eq.~(\ref{phi-psi}).  Using the boundary condition $(\partial_r \phi)_R=0$ and the relation~(\ref{psi}), we obtain
\be
W \equiv \int_{\Sigma} D_{\rm A} \, \boldsymbol{\nabla} c_{\rm A} \cdot d\boldsymbol{\Sigma} = \frac{1}{\ell} \int_{\Sigma} \left(\partial_r \psi \right)_R \, d\Sigma = - R \int_{\Sigma} \left(\partial_r Z \right)_R \, d\Sigma \, .
\ee
Next, the expansion~(\ref{Z-expansion}) shows that
\be
R\left(\partial_r Z\right)_R = - \sum_{lm} (l+1)\, z_{lm} \, Y_{lm} \, ,
\ee
leading to the result $W = \sqrt{4\pi} \, R^2 \, z_{00} $, 
since $d\Sigma = R^2 d\Omega$ with $d\Omega=\sin\theta d\theta d\varphi$ and $\int Y_{lm} \, d\Omega = \sqrt{4\pi} \, \delta_{l0} \, \delta_{m0}$.   Using Eq.~(\ref{z_l0}) for the coefficient $z_{00}$, the overall reaction rate of a single Janus particle is finally given by
\be\label{W-single-particle}
W= 4\pi\, R^2 \left( a_0 \, \varrho + \frac{3R}{2} \, b_0 \, {\bf g}_\varrho \cdot {\bf u} \right)
\ee
in terms of the reaction rate~(\ref{varrho}) and its gradient~(\ref{g_varrho}) \cite{GK20}.  The second term involving the gradients of the molecular concentrations is negligible in front of the first under the conditions $(R\Vert\boldsymbol{\nabla} c_\varkappa\Vert/c_\varkappa) \ll 1$, i.e., when the characteristic length scales of the gradients are larger than the radius $R$ of the Janus particles.  In the case where the gradients can be considered to be equal to zero (so that ${\bf g}_\varrho=0$), the expression~(\ref{W-single-particle}) for the overall reaction rate reduces to its first term $W= 4\pi R^2 a_0 \varrho$ \cite{GK18,GK19}.

\section{The vectorial spherical harmonics and their properties}
\label{app:VSH}

The vectorial spherical harmonics are functions of the polar and azimuthal angles $(\theta,\varphi)$.  They are defined as follows in terms of the spherical harmonics $Y_{lm}$ \cite{BEG85},
\bea
&& \bUps_{lm} \equiv Y_{lm} \, {\bf 1}_r \, , \label{VSH-Ups} \\
&& \bPsi_{lm} \equiv r \, \boldsymbol{\nabla} Y_{lm} \, , \label{VSH-Psi} \\
&& \bPhi_{lm} \equiv {\bf r} \times \boldsymbol{\nabla} Y_{lm} \, , \label{VSH-Phi}
\eea
for $l=0,1,2,\dots$ and $m=-l,-l+1,\dots,l-1,l$, where ${\bf 1}_r\equiv{\bf r}/r$.  In the orthonormal basis of the units vectors $\{{\bf 1}_r,{\bf 1}_\theta,{\bf 1}_\varphi\}$ associated with the spherical coordinates, the latter two vectorial spherical harmonics can be expressed as
\bea
&& \bPsi_{lm} = \partial_\theta Y_{lm} \, {\bf 1}_\theta + \frac{\partial_\varphi Y_{lm}}{\sin\theta} \, {\bf 1}_\varphi \, , \label{VSH-Psi-2} \\
&& \bPhi_{lm} = - \frac{\partial_\varphi Y_{lm}}{\sin\theta} \, {\bf 1}_\theta + \partial_\theta Y_{lm} \, {\bf 1}_\varphi \, , \label{VSH-Phi-2}
\eea
showing that the three kinds of vectorial spherical harmonics are orthogonal to each other,
\be
\bUps_{lm} \cdot \bPsi_{lm}  =  \bUps_{lm} \cdot \bPhi_{lm}  = \bPsi_{lm} \cdot \bPhi_{lm} = 0 \, .
\ee
Since $Y_{00}=1/\sqrt{4\pi}$, we have that $\bPsi_{00}=0$ and $\bPhi_{00}=0$.

Furthermore, the vectorial spherical harmonics satisfy the relations
\bea
&& \langle \bUps_{lm} \vert \bUps_{l'm'} \rangle = \delta_{ll'} \delta_{mm'} \, , \label{ortho-Ups}\\
&& \langle \bPsi_{lm} \vert \bPsi_{l'm'} \rangle = l(l+1) \, \delta_{ll'} \delta_{mm'} \, , \label{ortho-Psi}\\
&& \langle \bPhi_{lm} \vert \bPhi_{l'm'} \rangle = l(l+1) \, \delta_{ll'} \delta_{mm'} \, , \label{ortho-Phi}
\eea
where
\be
\langle {\bf X} \vert {\bf Y} \rangle \equiv \int {\bf X}^*\cdot{\bf Y} \, d\Omega \, .
\ee
The divergences, the rotationals, and the Laplacians of vectorial spherical harmonics multiplied by some radial function are given by the following formulas \cite{BEG85},
\bea
&& \boldsymbol{\nabla}\cdot\left[ f(r) \, \bUps_{lm}\right] = \frac{1}{r^2}\, \frac{d}{dr}\left[ r^2 f(r) \right] Y_{lm} \, , \label{div(f*Ups)} \\
&& \boldsymbol{\nabla}\cdot\left[ g(r) \, \bPsi_{lm}\right] = -\frac{l(l+1)}{r}\, g(r) \, Y_{lm} \, , \label{div(f*Psi)} \\
&& \boldsymbol{\nabla}\cdot\left[ h(r) \, \bPhi_{lm}\right] = 0 \, , \label{div(f*Phi)}
\eea
\bea
&& \boldsymbol{\nabla}\times\left[ f(r) \, \bUps_{lm}\right] = -\frac{f(r)}{r} \, \bPhi_{lm} \, , \label{rot(f*Ups)} \\
&& \boldsymbol{\nabla}\times\left[ g(r) \, \bPsi_{lm}\right] = \frac{1}{r}\, \frac{d}{dr}\left[ r \, g(r)\right] \, \bPhi_{lm} \, , \label{rot(f*Psi)} \\
&& \boldsymbol{\nabla}\times\left[ h(r) \, \bPhi_{lm}\right] = - \frac{l(l+1)}{r} \, h(r) \, \bUps_{lm} - \frac{1}{r} \, \frac{d}{dr}\left[ r \, h(r)\right] \bPsi_{lm} \, , \label{rot(f*Phi)}
\eea
\bea
&& \nabla^2\left[ f(r) \, \bUps_{lm}\right] = \left\{ \frac{1}{r} \, \frac{d^2}{dr^2}\left[r \, f(r)\right]-\frac{l(l+1)+2}{r^2} \, f(r) \right\} \bUps_{lm} + \frac{2}{r^2} \, f(r) \, \bPsi_{lm} \, , \label{Lapl(f*Ups)} \\
&& \nabla^2\left[ g(r) \, \bPsi_{lm}\right] = \left\{ \frac{1}{r} \, \frac{d^2}{dr^2}\left[r \, g(r)\right]-\frac{l(l+1)}{r^2} \, g(r) \right\} \bPsi_{lm} + \frac{2l(l+1)}{r^2} \, g(r) \, \bUps_{lm} \, , \label{Lapl(f*Psi)} \\
&& \nabla^2\left[ h(r) \, \bPhi_{lm}\right] = \left\{ \frac{1}{r} \, \frac{d^2}{dr^2}\left[r \, h(r)\right]-\frac{l(l+1)}{r^2} \, h(r) \right\} \bPhi_{lm} \, . \label{Lapl(f*Phi)}
\eea

\section{The stationary one-particle profile of the velocity}
\label{app:velocity}

The expansion~(\ref{velocity-VSH}) of the velocity field inside and around a single Janus particle into vectorial spherical harmonics is determined by using successively the expression~(\ref{velocity-solid}) for the velocity inside the particle, Stokes' equations~(\ref{Stokes-eqs}), the incompressibility condition~(\ref{div.v=0}), the boundary conditions~(\ref{bc-v-inf}) at large distances, the boundary condition~(\ref{bc-v-1}) at $r=R$ in the direction normal to the surface, and the boundary conditions~(\ref{bc-v-2}) at $r=R$ in the two directions tangential to the surface.  According to Eqs.~(\ref{ortho-Ups})-(\ref{ortho-Phi}), the radial functions of the expansion~(\ref{velocity-VSH}) are in principle given by
\bea
&& f_{lm}(r) = \int \bUps_{lm}^* \cdot {\bf v} \, d\Omega \, , \label{f-Ups-v}\\
&& g_{lm}(r) = \frac{1}{l(l+1)} \int \bPsi_{lm}^* \cdot {\bf v} \, d\Omega \, , \label{g-Psi-v}\\
&& h_{lm}(r) = \frac{1}{l(l+1)} \int \bPhi_{lm}^* \cdot {\bf v} \, d\Omega \, . \label{h-Phi-v}
\eea

\subsection{Velocity inside the solid particle}

Since the center of the particle is located at ${\bf r}={\bf R}=0$ and moves with the velocity ${\bf V}$, while the particle rotates with the angular velocity $\boldsymbol{\Omega}$, the velocity field~(\ref{velocity-solid}) inside the particle is given by
\be
{\bf v}_{\rm solid}= {\bf V} + \boldsymbol{\Omega} \times {\bf r}
\label{v_solid-single}
\ee
for $r=\Vert{\bf r}\Vert < R$.  Using Eqs.~(\ref{f-Ups-v})-(\ref{h-Phi-v}), this vector field can be expanded into vectorial spherical harmonics according to
\be
{\bf v}_{\rm solid}= \sum_{m=-1}^{+1} \left( V_{1m} \, \bUps_{1m} +V_{1m} \, \bPsi_{1m} - r \, \Omega_{1m} \, \bPhi_{1m} \right)
\label{v_solid-VSH}
\ee
with
\bea
&& V_{11} = \sqrt{\frac{2\pi}{3}} \left( - V_x + \imath \, V_y \right) , \qquad V_{10} = \sqrt{\frac{4\pi}{3}} \, V_z \, , \qquad V_{1,-1} = -V_{11}^* \, ;  \label{V_1m}\\
&& \Omega_{11} = \sqrt{\frac{2\pi}{3}} \left( - \Omega_x + \imath \, \Omega_y \right) , \qquad \Omega_{10} = \sqrt{\frac{4\pi}{3}} \, \Omega_z \, , \qquad \Omega_{1,-1} = -\Omega_{11}^* \, . \label{Omega_1m}
\eea
We note that the particle velocity $\bf V$ and its angular velocity $\boldsymbol{\Omega}$ will be determined here below by the force-free and torque-free conditions for the particle.

\subsection{Stokes' equations}

Inserting the expansion~(\ref{velocity-VSH}) into Stokes' equations~(\ref{Stokes-eqs}) and using Eqs.~(\ref{Lapl(f*Ups)})-(\ref{Lapl(f*Phi)}) together with the orthogonality of the vectorial spherical harmonics, we obtain the following equations for the corresponding radial functions,
\bea
&& \frac{1}{r} \, \frac{d^2}{dr^2}( r \, f_{lm} ) - \frac{l(l+1)+2}{r^2} \, f_{lm} + \frac{2 l(l+1)}{r^2} \, g_{lm} = \frac{1}{\eta_0} \, \frac{dp_{lm}}{dr} \, , \label{f_lm-eq} \\
&& \frac{1}{r} \, \frac{d^2}{dr^2}( r \, g_{lm} ) - \frac{l(l+1)}{r^2} \, g_{lm} + \frac{2}{r^2} \, f_{lm} = \frac{1}{\eta_0} \, \frac{p_{lm}}{r} \, , \label{g_lm-eq} \\
&& \frac{1}{r} \, \frac{d^2}{dr^2}( r \, h_{lm} ) - \frac{l(l+1)}{r^2} \, h_{lm} = 0 \, . \label{h_lm-eq}
\eea
The equations for $f_{lm}$ and $g_{lm}$ are coupled together, but they can be decoupled introducing the following functions,
\bea
&& \xi_{lm}(r) \equiv \frac{f_{lm}(r) + (l+1) \, g_{lm}(r)}{(2l+1) \, a_{lm}} \, , \label{xi_lm} \\
&& \zeta_{lm}(r) \equiv \frac{-f_{lm}(r) + l \, g_{lm}(r)}{(2l+1) \, b_{lm}} \, . \label{zeta_lm}
\eea
They obey linear second-order inhomogeneous ordinary differential equations,
\bea
&& \frac{d^2}{dr^2}(r \, \xi_{lm} ) - \frac{l(l-1)}{r} \, \xi_{lm} = r^l \, , \label{eq-xi_lm} \\
&& \frac{d^2}{dr^2}(r \, \zeta_{lm} ) - \frac{(l+1)(l+2)}{r} \, \zeta_{lm} = r^{-l-1} \, , \label{eq-zeta_lm}
\eea
the solutions of which are given by
\bea
&& \xi_{lm}(r) = \frac{r^{l+1}}{2(2l+1)} + \frac{1}{a_{lm}} \left( c_{lm}^{(\xi)} \, r^{l-1} + d_{lm}^{(\xi)} \, r^{-l} \right) , \label{sol-xi_lm} \\
&& \zeta_{lm}(r) = -\frac{r^{-l}}{2(2l+1)} - \frac{1}{b_{lm}} \left( c_{lm}^{(\zeta)} \, r^{l+1} + d_{lm}^{(\zeta)} \, r^{-l-2} \right) \label{sol-zeta_lm}
\eea
with some coefficients $c_{lm}^{(\xi)}$, $d_{lm}^{(\xi)}$, $c_{lm}^{(\zeta)}$, and $d_{lm}^{(\zeta)}$ to be determined.

Therefore, the corresponding radial functions have the following forms,
\bea
&& f_{lm}(r) = \left[ \frac{l \, a_{lm}}{2(2l+1)} + (l+1) \, c_{lm}^{(\zeta)}\right] r^{l+1} 
+\left[ \frac{(l+1)\, b_{lm}}{2(2l+1)} + l \, d_{lm}^{(\xi)}\right] r^{-l} 
+ l \, c_{lm}^{(\xi)} \, r^{l-1} + (l+1) \, d_{lm}^{(\zeta)} \, r^{-l-2} \, , \qquad \label{sol-f_lm} \\
&& g_{lm}(r) = \left[ \frac{a_{lm}}{2(2l+1)} - c_{lm}^{(\zeta)}\right] r^{l+1} 
+ \left[-\frac{b_{lm}}{2(2l+1)} + d_{lm}^{(\xi)}\right] r^{-l} 
+ c_{lm}^{(\xi)} \, r^{l-1} - d_{lm}^{(\zeta)} \, r^{-l-2} \,  . \label{sol-g_lm}
\eea

Furthermore, Eq.~(\ref{h_lm-eq}) implies that 
\be
h_{lm}(r) = x_{lm} \, r^l + y_{lm} \, r^{-l-1}
\label{sol-h_lm}
\ee
with some coefficients $x_{lm}$ and $y_{lm}$ to be determined.

We note that the functions $g_{00}(r)$ and $h_{00}(r)$ are not needed since $\bPsi_{00}=0$ and $\bPhi_{00}=0$.

\subsection{The incompressibility condition}

Inserting the expansion~(\ref{velocity-VSH}) into the incompressibility condition~(\ref{div.v=0}) and using Eqs.~(\ref{div(f*Ups)})-(\ref{div(f*Phi)}) together with the orthogonality of the vectorial spherical harmonics, we get
\be
\boldsymbol{\nabla}\cdot{\bf v} = \sum_{lm} \left[ \frac{1}{r^2} \, \frac{d}{dr}\left( r^2 \, f_{lm} \right) - \frac{l(l+1)}{r} \, g_{lm} \right] Y_{lm} = 0 \, ,
\ee
so that the radial functions $f_{lm}$ and $g_{lm}$ should be related by
\be
\frac{d}{dr}\left( r^2 \, f_{lm} \right) = l(l+1) \, r \, g_{lm} \, .
\ee
Inserting therein the solutions~(\ref{sol-f_lm}) and~(\ref{sol-g_lm}), we find that the two following series of coefficients are determined
\bea
&& c_{lm}^{(\zeta)} = - \frac{l \, a_{lm}}{(l+1)(2l+3)(2l+1)} \, , \label{c_lm(zeta)} \\
&& d_{lm}^{(\xi)} = \frac{(l+1) \, b_{lm}}{l(2l-1)(2l+1)} \, . \label{d_lm(xi)}
\eea
For $l=0$, the incompressibility condition implies that $c_{00}^{(\zeta)}=0$ and $b_{00}=0$.
As a consequence of incompressibility, the functions~(\ref{sol-f_lm}) and~(\ref{sol-g_lm}) thus become
\bea
&& f_{lm}(r) = \frac{l \, a_{lm}}{2(2l+3)} \, r^{l+1} 
+ \frac{(l+1)\, b_{lm}}{2(2l-1)} \, r^{-l} 
+ l \, c_{lm}^{(\xi)} \, r^{l-1} + (l+1) \, d_{lm}^{(\zeta)} \, r^{-l-2}
\qquad\quad \mbox{if} \quad l\ge 0 \, , \label{sol-f_lm-2} \\
&& g_{lm}(r) = \frac{(l+3)\, a_{lm}}{2(l+1)(2l+3)} \, r^{l+1} 
+ \frac{(2-l)\, b_{lm}}{2l(2l-1)} \, r^{-l} 
+ c_{lm}^{(\xi)} \, r^{l-1} - d_{lm}^{(\zeta)} \, r^{-l-2} 
\qquad\quad\ \ \mbox{if} \quad l\ge 1 \, . \label{sol-g_lm-2}
\eea
If $l=0$, Eq.~(\ref{sol-f_lm-2}) gives $f_{00}(r) = \, d_{00}^{(\zeta)} r^{-2}$, because $b_{00}=0$.  Therefore, the coefficients $a_{00}$ and $c_{00}^{(\xi)}$ do not appear in the problem and they can be discarded.

\subsection{The velocity field far from the particle}

At large distances from the particle, the boundary conditions~(\ref{bc-v-inf}) should be satisfied.  According to Eq.~(\ref{asympt-v}), the velocity field should thus behave as ${\bf v} = {\bf v}^{\circ} + {\boldsymbol{\mathsf G}}\cdot{\bf r}$ for $r\to\infty$, i.e., it may not grow faster than the first power of the radial distance $r$.  According to Eqs.~(\ref{sol-f_lm-2}) and~(\ref{sol-g_lm-2}), all the coefficients $a_{lm}$ with $l\ge 1$ and all the coefficients $c_{lm}^{(\xi)}$ with $l\ge 3$ should thus be equal to zero.  Furthermore, according to Eq.~(\ref{sol-h_lm}), all the coefficients $x_{lm}$ with $l\ge 2$ should also be equal to zero.  Out of the coefficients $c_{lm}^{(\xi)}$ and $x_{lm}$, there remain $c_{1m}^{(\xi)}$, $c_{2m}^{(\xi)}$, and $x_{1m}$ to be determined by the constant velocity ${\bf v}^{\circ}$ and the velocity gradients ${\boldsymbol{\mathsf G}}$.  Using Eqs.~(\ref{f-Ups-v})-(\ref{h-Phi-v}), we find that 
\be
{\bf v}^{\circ} + {\boldsymbol{\mathsf G}}\cdot{\bf r} = \sum_{m=-1}^{+1} c_{1m}^{(\xi)} \left( \bUps_{1m} +\bPsi_{1m}\right) + r \sum_{m=-2}^{+2} \left( 2\, c_{2m}^{(\xi)} \, \bUps_{2m} + c_{2m}^{(\xi)} \, \bPsi_{2m} + x_{1m}\,  \bPhi_{1m} \right)
\label{asymptotic-velocity}
\ee
with
\bea
&& c_{11}^{(\xi)} = \sqrt{\frac{2\pi}{3}} \left( - v_x^{\circ} + \imath \, v_y^{\circ} \right) , \qquad c_{10}^{(\xi)} = \sqrt{\frac{4\pi}{3}} \, v_z^{\circ} \, , \qquad c_{1,-1}^{(\xi)} = -c_{11}^{(\xi)*} \, ;  \label{c_1m^xi}\\
&& c_{22}^{(\xi)} = \sqrt{\frac{\pi}{30}} \left[ G_{xx}-G_{yy} - \imath \left( G_{xy}+G_{yx} \right)\right] , \qquad 
c_{21}^{(\xi)} = \sqrt{\frac{\pi}{30}} \left[-G_{xz}-G_{zx} + \imath \left( G_{yz}+G_{zy} \right)\right] , \nonumber\\
&& c_{20}^{(\xi)} = -\sqrt{\frac{\pi}{5}} \left( G_{xx} + G_{yy} \right) , \qquad
 c_{2,-1}^{(\xi)} = -c_{21}^{(\xi)*} \, , \qquad
 c_{2,-2}^{(\xi)} = c_{22}^{(\xi)*} \, ;  \label{c_2m^xi}\\
&& x_{11} = \sqrt{\frac{\pi}{6}} \left[-G_{yz} + G_{zy} - \imath \left( G_{xz} - G_{zx} \right)\right] , \qquad
x_{10} = \sqrt{\frac{\pi}{3}} \left( G_{xy} - G_{yx} \right) , \qquad
x_{1,-1} = - x_{11}^* \, .  \label{x_1m}
\eea

We note that the vorticity at large distances from the particle is given by
\be
\boldsymbol{\omega}_\infty = (\boldsymbol{\nabla}\times{\bf v})_\infty=(G_{zy}-G_{yz})\, {\bf 1}_x + (G_{xz}-G_{zx})\, {\bf 1}_y + (G_{yx}-G_{xy})\, {\bf 1}_z = - 2 \sum_{m=-1}^{+1} x_{1m} \left(\bUps_{1m}+\bPsi_{1m}\right) .
\label{asymptotic-omega}
\ee
Therefore, the coefficients $x_{1m}$ determine the vorticity far from the particle according to $x_{1m}=-(\boldsymbol{\omega}_\infty)_{1m}/2$.

\subsection{The boundary condition normal to the particle surface}

Now, we consider the boundary condition~(\ref{bc-v-1}) for a single particle located at the origin ${\bf r}=0$ of the reference frame, which reads
\be
\left({\bf n}\cdot{\bf v}\right)_R = {\bf n}\cdot{\bf V} \, , 
\label{bc-v-1-single}
\ee
because the velocity field of the solid particle is given by Eq.~(\ref{v_solid-single}) and ${\bf n}={\bf r}/r$ is orthogonal to $\boldsymbol{\Omega}\times{\bf r}$.  The left-hand side of this boundary condition is the radial component $\left({\bf n}\cdot{\bf v}\right)_R =\left({\bf 1}_r\cdot{\bf v}\right)_R =(v_r)_R$ of the velocity field~(\ref{velocity-VSH}) at $r=R$, since ${\bf n}={\bf 1}_r$.  Its right-hand side is the radial component of the particle velocity, ${\bf n}\cdot{\bf V} ={\bf 1}_r\cdot{\bf V} =V_r$.  Using the expansions~(\ref{velocity-VSH}) and~(\ref{v_solid-VSH}) with the coefficients~(\ref{V_1m}), the boundary condition~(\ref{bc-v-1-single}) gives
\be
(v_r)_R=\sum_{lm} f_{lm}(R) \, Y_{lm} = \sum_{m=-1}^{+1} V_{1m} \, Y_{1m} \, .
\label{bc-v-1-single-SH}
\ee
Using Eq.~(\ref{sol-f_lm-2}) with the already known coefficients $b_{00}=0$, $a_{lm}=0$ for $l\ge 1$, and $c_{lm}^{(\xi)}=0$ for $l\ge 3$, the boundary condition~(\ref{bc-v-1-single-SH}) is satisfied if
\bea
&& f_{00}(R) = d_{00}^{(\zeta)} \, R^{-2} = 0 \, , \\
&& f_{1m}(R) = b_{1m} \, R^{-1} + c_{1m}^{(\xi)} + 2 \, d_{1m}^{(\zeta)} \, R^{-3} = V_{1m} \, , \\
&& f_{2m}(R) = \frac{b_{2m}}{2} \, R^{-2} + 2 \, c_{2m}^{(\xi)} \, R + 3 \, d_{2m}^{(\zeta)} \, R^{-4}  = 0 \, , \\
&& f_{lm}(R) = \frac{(l+1)\, b_{lm}}{2(2l-1)} \, R^{-l} + (l+1) \, d_{lm}^{(\zeta)} \, R^{-l-2} = 0
\qquad\quad \mbox{for} \ \ l\ge 3 \, ,
\eea
from which we deduce that
\bea
&& d_{00}^{(\zeta)} = 0 \, , \\
&& d_{1m}^{(\zeta)} = \frac{R^3}{2} \left( V_{1m} - c_{1m}^{(\xi)} - \frac{b_{1m}}{R} \right) \, , \\
&& d_{2m}^{(\zeta)} = \frac{R^4}{3} \left( - 2 \, c_{2m}^{(\xi)} \, R - \frac{b_{2m}}{2 R^2}\right) \, , \\
&& d_{lm}^{(\zeta)}  =  - \frac{b_{lm} R^2}{2(2l-1)} \qquad\quad \mbox{for} \ \ l\ge 3 \, .
\eea
Hence, the radial functions of the velocity field become
\bea
&& f_{00}(r) = 0 \, , \label{f_00}\\
&& f_{1m}(r) = V_{1m} \, \frac{R^3}{r^3} + c_{1m}^{(\xi)} \left( 1 - \frac{R^3}{r^3} \right) + \frac{b_{1m}}{r} \left( 1 - \frac{R^2}{r^2} \right) \, , \label{f_1m}\\
&& f_{2m}(r) = 2 c_{2m}^{(\xi)}\, r \left( 1 - \frac{R^5}{r^5} \right) + \frac{b_{2m}}{2\, r^2 } \left( 1 - \frac{R^2}{r^2} \right) \, , \label{f_2m}\\
&& f_{lm}(r) = \frac{(l+1) b_{lm}}{2(2l-1)\, r^l} \left( 1 - \frac{R^2}{r^2} \right)
\qquad\quad \mbox{for} \ \ l\ge 3 \, , \label{f_lm}
\eea
\bea
&& g_{1m}(r) = -\frac{V_{1m}}{2} \, \frac{R^3}{r^3} + c_{1m}^{(\xi)} \left( 1 + \frac{R^3}{2\, r^3} \right) + \frac{b_{1m}}{2 \, r} \left( 1 + \frac{R^2}{r^2} \right) \, , \label{g_1m}\\
&& g_{2m}(r) = c_{2m}^{(\xi)}\, r \left( 1 + \frac{2\, R^5}{3\, r^5} \right) + \frac{b_{2m}\, R^2}{6\, r^4} \, , \label{g_2m}\\
&& g_{lm}(r) = \frac{b_{lm}}{2(2l-1)\, r^l} \left(-\frac{l-2}{l} + \frac{R^2}{r^2} \right)
\qquad\quad \mbox{for} \ \ l\ge 3 \, , \label{g_lm}
\eea
\bea
&& h_{1m}(r) = x_{1m} \, r + y_{1m} \, r^{-2}  \, , \label{h_1m}\\
&& h_{2m}(r) = y_{2m} \, r^{-3} \, , \label{h_2m}\\
&& h_{lm}(r) = y_{lm} \, r^{-l-1}  
\qquad\quad \mbox{for} \ \ l\ge 3 \, , \label{h_lm}
\eea
We note that the components of the particle velocity always appear in the combination $V_{1m}-c_{1m}^{(\xi)}$  corresponding to the difference ${\bf V}-{\bf v}^{\circ}$ with respect to the constant velocity of the fluid, which is a consequence of Galilean invariance.

The coefficients $c_{1m}^{(\xi)}$, $c_{2m}^{(\xi)}$, and $x_{1m}$ are given by Eqs.~(\ref{c_1m^xi}), (\ref{c_2m^xi}), and~(\ref{x_1m}). Thus, the coefficients $V_{1m}$, $b_{lm}$, and $y_{lm}$ should still be determined.

\subsection{The boundary conditions tangential to the particle surface}

In the two directions tangential to the interface between the particle and the fluid, the velocity field should satisfy the boundary conditions~(\ref{bc-v-2}).  For a single particle located at ${\bf r}=0$, they read
\be
\left({\boldsymbol{\mathsf 1}}-{\bf n}{\bf n}\right) \cdot \left[{\bf v} -{\bf v}_{\rm solid}
- b\left(\pmb{\nabla}{\bf v}+\pmb{\nabla}{\bf v}^{\rm T}\right)\cdot{\bf n} +\sum_\varkappa b_\varkappa \boldsymbol{\nabla}c_\varkappa\right]_R = 0 \, ,
\label{bc-v-2-single}
\ee
where the velocity of the solid particle is given by Eq.~(\ref{v_solid-single}) and the diffusiophoretic terms coupling the velocity with the molecular concentrations by Eq.~(\ref{sum-bk-gk}), the function $Z(r,\theta,\varphi)$ being expressed in terms of the coefficients~(\ref{a_l-b_l-c_l})-(\ref{M-N}) by Eqs.~(\ref{Z-expansion})-(\ref{z_lm}) or, equivalently, by Eqs.~(\ref{Z-K-L})-(\ref{Lx-Ly}).

Since ${\bf n}={\bf 1}_r$, we have that ${\boldsymbol{\mathsf 1}}-{\bf n}{\bf n}={\bf 1}_\theta {\bf 1}_\theta + {\bf 1}_\varphi {\bf 1}_\varphi $, which is the projection onto the plane tangential to the particle interface.  All the vectors can be decomposed in the orthonormal basis $\{ {\bf 1}_r, {\bf 1}_\theta , {\bf 1}_\varphi \}$.  In particular, the velocity field can be equivalently expressed as ${\bf v} = v_r{\bf 1}_r+ v_\theta {\bf 1}_\theta + v_\varphi {\bf 1}_\varphi $.  In the two tangential directions, the boundary conditions~(\ref{bc-v-2-single}) give
\bea
&& \left(v_\theta \right)_R = {\bf 1}_\theta \cdot \left( {\bf V} + \boldsymbol{\Omega}\times{\bf r}\right)_R +  b \, {\bf 1}_\theta \cdot \left(\pmb{\nabla}{\bf v}+\pmb{\nabla}{\bf v}^{\rm T}\right)_R\cdot{\bf 1}_r - {\bf 1}_\theta \cdot \left(\sum_\varkappa b_\varkappa \boldsymbol{\nabla}c_\varkappa\right)_R , \\
&& \left(v_\varphi \right)_R = {\bf 1}_\varphi \cdot \left( {\bf V} + \boldsymbol{\Omega}\times{\bf r}\right)_R +  b \, {\bf 1}_\varphi \cdot \left(\pmb{\nabla}{\bf v}+\pmb{\nabla}{\bf v}^{\rm T}\right)_R\cdot{\bf 1}_r - {\bf 1}_\varphi \cdot \left(\sum_\varkappa b_\varkappa \boldsymbol{\nabla}c_\varkappa\right)_R .
\eea

For the velocity of the solid particle, we have that
\be
\left({\boldsymbol{\mathsf 1}}-{\bf n}{\bf n}\right) \cdot \left( {\bf V} + \boldsymbol{\Omega}\times{\bf r}\right)_R = \left(V_\theta + R \, \Omega_\varphi \right) {\bf 1}_\theta + \left(V_\varphi - R \, \Omega_\theta \right) {\bf 1}_\varphi = \sum_{lm} \left( V_{1m} \, \delta_{l1} \, \bPsi_{lm} - R \, \Omega_{1m} \, \delta_{l1} \, \bPhi_{lm} \right) ,
\ee
because of Eq.~(\ref{v_solid-VSH}).

According to Eq.~(15.20) of Ref.~\cite{LL59} and using the expansion~(\ref{velocity-VSH}), the components of the symmetrised tensor of velocity gradients are given by
\bea
&& {\bf 1}_\theta \cdot \left(\pmb{\nabla}{\bf v}+\pmb{\nabla}{\bf v}^{\rm T}\right)_R\cdot{\bf 1}_r  = \left( \partial_r v_\theta + \frac{\partial_\theta v_r}{r} - \frac{v_\theta}{r} \right)_R = \sum_{lm} \left( G_{lm} \, \partial_\theta Y_{lm} - H_{lm} \, \frac{\partial_\varphi Y_{lm}}{\sin\theta} \right) , \\
&& {\bf 1}_\varphi \cdot \left(\pmb{\nabla}{\bf v}+\pmb{\nabla}{\bf v}^{\rm T}\right)_R\cdot{\bf 1}_r = \left( \partial_r v_\varphi + \frac{\partial_\varphi v_r}{r \sin\theta} - \frac{v_\varphi}{r} \right)_R = \sum_{lm} \left( H_{lm} \, \partial_\theta Y_{lm} + G_{lm} \, \frac{\partial_\varphi Y_{lm}}{\sin\theta} \right) ,
\eea
where the coefficients
\be
G_{lm} \equiv \left(\frac{dg_{lm}}{dr} - \frac{g_{lm}}{r} + \frac{f_{lm}}{r}\right)_R
\qquad\mbox{and}\qquad
H_{lm} \equiv \left(\frac{dh_{lm}}{dr} - \frac{h_{lm}}{r} \right)_R
\ee
take the following values,
\bea
&& G_{1m} = \frac{3}{R} \left( V_{1m} - c_{1m}^{(\xi)} \right) - \frac{3}{R^2} \, b_{1m} \, , \\
&& G_{2m} = -\frac{10}{3} \, c_{2m}^{(\xi)} - \frac{5}{6\, R^3} \, b_{2m} \, , \\
&& G_{lm} = - \frac{2l+1}{l(2l-1)\, R^{l+1}} \, b_{lm} 
\qquad\mbox{for} \quad l \ge 3 \, , \\
&& H_{1m} = - \frac{3}{R^3} \, y_{1m} \, , \\
&& H_{lm} = - \frac{l+2}{R^{l+2}} \, y_{lm} \qquad\mbox{for} \quad l \ge 2 \, .
\eea
Consequently, we have that
\be
\left({\boldsymbol{\mathsf 1}}-{\bf n}{\bf n}\right) \cdot \left(\pmb{\nabla}{\bf v}+\pmb{\nabla}{\bf v}^{\rm T}\right)\cdot{\bf 1}_r = \sum_{lm} \left( G_{lm} \, \bPsi_{lm} + H_{lm} \, \bPhi_{lm} \right) .
\ee

Taking the scalar product of the boundary conditions~(\ref{bc-v-2-single}) with the vectorial spherical harmonics $\bPsi_{lm}$ and $\bPhi_{lm}$ and using the relations~(\ref{ortho-Psi}) and~(\ref{ortho-Phi}), we infer that
\bea
&& g_{lm}(R) = V_{1m} \, \delta_{l1} + b \, G_{lm} - X_{lm}^{(g)}
\qquad\qquad\mbox{with}\qquad
X_{lm}^{(g)} = \frac{1}{l(l+1)} \int d\Omega \, \bPsi_{lm}^* \cdot  \left(\sum_\varkappa b_\varkappa \boldsymbol{\nabla}c_\varkappa\right)_R , \label{X_lm^g}\\
&& h_{lm}(R) = - R \, \Omega_{1m} \, \delta_{l1} + b \, H_{lm} - X_{lm}^{(h)}
\qquad\mbox{with}\qquad
X_{lm}^{(h)} = \frac{1}{l(l+1)} \int d\Omega \, \bPhi_{lm}^* \cdot  \left(\sum_\varkappa b_\varkappa \boldsymbol{\nabla}c_\varkappa\right)_R . \label{X_lm^h}
\eea
As a consequence, we find that
\bea
&& b_{1m} = - \frac{R}{\displaystyle 1 + \frac{3b}{R}} \left[ \frac{3}{2} \left( 1 + \frac{2b}{R}\right) \left( c_{1m}^{(\xi)} - V_{1m} \right) + X_{1m}^{(g)} \right] , \label{b_1m-fin}\\
&& b_{2m} = - \frac{6\, R^2}{\displaystyle 1 + \frac{5b}{R}} \left[ \frac{5}{3} \left( 1 + \frac{2b}{R}\right) c_{2m}^{(\xi)} \, R + X_{2m}^{(g)} \right] , \label{b_2m-fin}\\
&& b_{lm} = - \frac{l(2l-1)\, R^l}{\displaystyle 1 + \frac{(2l+1)b}{R}} \, X_{lm}^{(g)} \qquad\mbox{for} \quad l \ge 3 \,  , \label{b_lm-fin}\\
&& y_{1m} = - \frac{R^2}{\displaystyle 1 + \frac{3b}{R}} \left[ R \left( x_{1m} + \Omega_{1m} \right) + X_{1m}^{(h)}\right] , \label{y_1m-fin}\\
&& y_{lm} = - \frac{R^{l+1}}{\displaystyle 1 + \frac{(l+2)b}{R}} \, X_{lm}^{(h)} \qquad\mbox{for} \quad l \ge 2 \, . \label{y_lm-fin}
\eea

In addition to the calculation of the coefficients~(\ref{X_lm^g}) and~(\ref{X_lm^h}), there remains to determine the consequences of the force-free and torque-free conditions.

\subsection{The force-free condition}

Using the results of Refs.~\cite{MB74,ABM75}, the force~(\ref{force}) exerted by the fluid on a single Janus particle was shown in Refs.~\cite{GK18,GK19} to be given by
\be
{\bf F} = - \frac{6\pi\eta_0 R}{\displaystyle 1 + \frac{3b}{R}} \left[\left( 1 +\frac{2b}{R}\right) \left({\bf V}-{\bf v}^{\circ}\right) -   \sum_\varkappa \overline{b_\varkappa \left({\boldsymbol{\mathsf 1}}-{\bf n}{\bf n}\right) \cdot \boldsymbol{\nabla}c_\varkappa}^{\rm s} \right] ,
\label{force-single}
\ee
where
\be
\overline{(\cdot)}^{\rm s} = \frac{1}{4\pi} \int (\cdot)_R \, d\Omega
\ee
denotes the average over the spherical surface of the particle. In particular, the surface averages of the vectorial spherical harmonics we here need can be calculated, leading to
\be
\overline{\bPsi_{1m}}^{\rm s} = \frac{2}{3} \left( \bUps_{1m} + \bPsi_{1m} \right), \qquad
\overline{\bPsi_{lm}}^{\rm s} = 0 \qquad\mbox{for} \quad l \ge 2 \, , 
\qquad\mbox{and}\qquad
\overline{\bPhi_{lm}}^{\rm s} = 0 \qquad\mbox{for} \quad l \ge 1 \, .
\label{VSH-surf-averages}
\ee
Since
\be
\left[\sum_\varkappa b_\varkappa \left({\boldsymbol{\mathsf 1}}-{\bf n}{\bf n}\right) \cdot \boldsymbol{\nabla}c_\varkappa\right]_R = \sum_{lm} \left( X_{lm}^{(g)} \, \bPsi_{lm} + X_{lm}^{(h)} \, \bPhi_{lm} \right)
\qquad\mbox{with} \quad l \ge 1 \, ,
\ee
we have that
\be
\sum_\varkappa \overline{ b_\varkappa \left({\boldsymbol{\mathsf 1}}-{\bf n}{\bf n}\right) \cdot \boldsymbol{\nabla}c_\varkappa}^{\rm s} = \frac{2}{3} \sum_{m=-1}^{+1} X_{1m}^{(g)} \left( \bUps_{1m} + \bPsi_{1m} \right) .
\ee
Now, the particle velocity can be expanded as ${\bf V} = \sum_{m=-1}^{+1} V_{1m} \left( \bUps_{1m} + \bPsi_{1m} \right)$ according to Eqs.~(\ref{v_solid-single}) and~(\ref{v_solid-VSH}) with $\boldsymbol{\Omega}=0$.  In addition, the constant velocity has a similar form ${\bf v}^{\circ} = \sum_{m=-1}^{+1} c_{1m}^{(\xi)} \left( \bUps_{1m} + \bPsi_{1m} \right)$ because of Eq.~(\ref{asymptotic-velocity}) with $\boldsymbol{\mathsf G} = 0$.
Therefore, the force~(\ref{force-single}) can be equivalently expressed as
\be
{\bf F} = -4\pi \eta_0 \sum_{m=-1}^{+1} b_{1m} \left( \bUps_{1m} + \bPsi_{1m} \right)
\ee
in terms of the coefficients~(\ref{b_1m-fin}).

The result is that the force-free condition ${\bf F}=0$ is equivalent to the condition $b_{1m}=0$.

\subsection{The torque-free condition}

The torque~(\ref{torque}) exerted by the fluid on the Janus particle was evaluated in Refs.~\cite{GK18,GK19} with the methods of Refs.~\cite{H75,F76a,F76b} and is given by
\be
{\bf T} = - \frac{8\pi\eta_0 R^3}{\displaystyle 1 + \frac{3b}{R}} \left[\left(\boldsymbol{\Omega} - \frac{1}{2}\, \boldsymbol{\omega}_\infty \right) -  \frac{3}{2 R^2}  \sum_\varkappa \overline{b_\varkappa {\bf r}\times \boldsymbol{\nabla}c_\varkappa}^{\rm s} \right] ,
\label{torque-single}
\ee
where $\boldsymbol{\omega}_\infty=(\boldsymbol{\nabla}\times{\bf v})_\infty$ is the vorticity at large distances from the particle.
Here, we have that 
\be
\frac{1}{R}\left(\sum_\varkappa b_\varkappa {\bf r}\times  \boldsymbol{\nabla}c_\varkappa\right)_R = \sum_{lm} \left( X_{lm}^{(g)} \, \bPhi_{lm} - X_{lm}^{(h)} \, \bPsi_{lm} \right)
\qquad\mbox{with} \quad l \ge 1 \, ,
\ee
so that
\be
\frac{1}{R}\sum_\varkappa \overline{b_\varkappa {\bf r}\times  \boldsymbol{\nabla}c_\varkappa}^{\rm s} = - \frac{2}{3}\sum_{m=-1}^{+1} X_{1m}^{(h)} \left( \bUps_{1m} + \bPsi_{1m} \right) \, ,
\ee
as a consequence of Eq.~(\ref{VSH-surf-averages}).

Since
\be
\boldsymbol{\Omega} - \frac{1}{2} \, \boldsymbol{\omega}_\infty = \sum_{m=-1}^{+1} \left[\Omega_{1m} - \frac{1}{2} \, (\boldsymbol{\omega}_\infty)_{1m}\right] \left( \bUps_{1m} + \bPsi_{1m} \right) = \sum_{m=-1}^{+1} \left(\Omega_{1m} + x_{1m}\right) \left( \bUps_{1m} + \bPsi_{1m} \right)
\ee
according to Eq.~(\ref{asymptotic-omega}),
the torque~(\ref{torque-single}) has the following equivalent expression,
\be
{\bf T} = 8\pi \eta_0 \sum_{m=-1}^{+1} y_{1m} \left( \bUps_{1m} + \bPsi_{1m} \right)
\ee
in terms of the coefficients~(\ref{y_1m-fin}).

Here, the result is that the torque-free condition ${\bf T}=0$ is equivalent to the condition $y_{1m}=0$.

\subsection{The asymptotic velocity and pressure}

Since all the coefficients are now determined, we can consider the asymptotic expressions for the velocity~(\ref{velocity-VSH}) and pressure~(\ref{pressure-SH}) around a single Janus particle.
After having taken into account the boundary conditions on the surface of the particle and at infinity, we have obtained the radial functions~(\ref{f_00})-(\ref{h_lm}) with the coefficients~(\ref{b_1m-fin})-(\ref{y_lm-fin}), where $b_{1m}=0$ and $y_{1m}=0$ because of the force-free and torque-free conditions.  Therefore, the velocity and pressure fields can be expressed as supposed in Eqs.~(\ref{asympt-v}) and~(\ref{asympt-p}), where the disturbance velocity and pressure behave according to Eqs.~(\ref{Dv-O}) and~(\ref{Dp-O}) for $r\to\infty$.  

On the one hand, the background velocity has the form given by Eq.~(\ref{asymptotic-velocity}) as confirmed by the terms going as $c_{1m}^{(\xi)}$ in Eqs.~(\ref{f_1m}) and~(\ref{g_1m}), as $c_{2m}^{(\xi)}\, r$ in Eqs.~(\ref{f_2m}) and~(\ref{g_2m}), and as $x_{1m}^{(\xi)}\, r$ in Eq.~(\ref{h_1m}).  Furthermore, the leading term of the disturbance velocity $\Delta{\bf v}$ goes as $1/r^2$.  Such a dependence on $r$ is found in Eq.~(\ref{f_2m}) with the coefficient $b_{2m}$ given by Eq.~(\ref{b_2m-fin}).  There is another such dependence in the radial function~(\ref{h_1m}), but its coefficient is $y_{1m}=0$ because of the torque-free condition.  Accordingly, the disturbance velocity has the following asymptotic profile,
\be
\Delta{\bf v} = -\frac{R^2}{r^2} \, \frac{1}{\displaystyle 1 + \frac{5b}{R}} \sum_{m=-2}^{+2} \left[ 5 \left( 1 + \frac{2b}{R}\right) c_{2m}^{(\xi)} \, R + 3 \, X_{2m}^{(g)} \right] \bUps_{2m} + O\left(\frac{1}{r^3}\right) ,
\label{Dv-final}
\ee
where the coefficients $X_{2m}^{(g)}$ are given by Eq.~(\ref{X_lm^g}) with $l=2$ and the coefficients $c_{2m}^{(\xi)}$ by Eqs.~(\ref{c_2m^xi}).

On the other hand, the pressure is given by the expansion~(\ref{pressure-SH}) into spherical harmonics with the radial functions~(\ref{p_lm}), where $a_{lm}=0$ for $l\ge 1$ to obtain the asymptotic velocity~(\ref{asymptotic-velocity}), $b_{00}=0$ to satisfy the incompressibility condition~(\ref{d_lm(xi)}) for $l=0$, and $b_{1m}=0$ for the force-free condition to hold.  In Eq.~(\ref{pressure-SH}), the background value of the pressure is given by $p^{\circ}=p_{00} Y_{00}=\eta_0 a_{00} Y_{00}$, since $a_{00}$ is the coefficient of the constant profile going as $r^0$ in Eq.~(\ref{p_lm}) with $l=0$ and because it can take an arbitrary value.  Now, Eq.~(\ref{b_2m-fin}) implies that the disturbance pressure has the following asymptotic expression,
\be
\Delta p = -\frac{R^2}{r^3} \, \frac{2\, \eta_0}{\displaystyle 1 + \frac{5b}{R}} \sum_{m=-2}^{+2} \left[ 5 \left( 1 + \frac{2b}{R}\right) c_{2m}^{(\xi)} \, R + 3 \, X_{2m}^{(g)} \right] Y_{2m} + O\left(\frac{1}{r^4}\right) .
\label{Dp-final}
\ee
The asymptotic profiles~(\ref{Dv-final}) and~(\ref{Dp-final}) are consistent with the expected behaviours~(\ref{Dv-O}) and~(\ref{Dp-O}) as a consequence of the force-free and torque-free conditions.  These expressions are thus suitable to evaluate the effect due to the presence of Janus particles on the stress tensor using Eq.~(\ref{Pij-suspension-LL}).

According to the formula~(\ref{sum-bk-gk}), the coefficients $X_{2m}^{(g)}$ introduced in Eq.~(\ref{X_lm^g}) with $l=2$ can be similarly decomposed as
\be
X_{2m}^{(g)} = X_{2m}^{(g,{\rm d})} + X_{2m}^{(g,{\rm r})} 
\ee
into a contribution from simple diffusiophoresis and another one from the reaction,
\bea
&& X_{2m}^{(g,{\rm d})} \equiv \frac{1}{4} \int d\Omega \, \bPsi_{2m}^* \cdot \left( b_{\rm A} {\bf g}_{\rm A} + b_{\rm B} {\bf g}_{\rm B}\right) , \label{X_2m^gd} \\
&& X_{2m}^{(g,{\rm r})}  \equiv \frac{R}{6} \int d\Omega \,  \bPsi_{2m}^* \cdot \left[\left(\frac{b_{\rm B}}{D_{\rm B}} - \frac{b_{\rm A}}{D_{\rm A}}\right)(\boldsymbol{\nabla} Z)_R\right] .
\label{X_2m^gr}
\eea
We note that, in Eq.~(\ref{X_2m^gd}), the projector $\left({\boldsymbol{\mathsf 1}}-{\bf n}{\bf n}\right)$ has been eliminated because the vectorial spherical harmonics $\bPsi_{2m}$ are oriented tangentially to the particle surface, i.e., inside the subspace of the projector.

In terms of the coefficients $b_{2m}$ given by Eq.~(\ref{b_2m-fin}), Eqs.~(\ref{Dv-final}) and~(\ref{Dp-final}) read
\be
\Delta{\bf v} = \frac{1}{2\, r^2} \sum_{m=-2}^{+2} b_{2m} \, Y_{2m} \, {\bf 1}_r + O\left(\frac{1}{r^3}\right)
\qquad\mbox{and}\qquad
\Delta p = \frac{\eta_0}{r^3} \sum_{m=-2}^{+2} b_{2m} \, Y_{2m} + O\left(\frac{1}{r^4}\right) ,
\label{Dv-Dp-b_2m}
\ee
where the coefficients $b_{2m}$ in Eq.~(\ref{Dv-Dp-b_2m}) can be decomposed as
\be
b_{2m} = b_{2m}^{({\rm v})} + b_{2m}^{({\rm d})} + b_{2m}^{({\rm r})}
\ee
into the three following contributions,
\bea
&& b_{2m}^{({\rm v})}  \equiv -10 \, R^3 \, \frac{\displaystyle 1 + \frac{2b}{R}}{\displaystyle 1 + \frac{5b}{R}} \, c_{2m}^{(\xi)} \, , \label{b_2m^v} \\
&& b_{2m}^{({\rm d})} \equiv -\frac{3}{2}\, \frac{R^2}{\displaystyle 1 + \frac{5b}{R}} \int d\Omega \, \bPsi_{2m}^* \cdot \left( b_{\rm A} {\bf g}_{\rm A} + b_{\rm B} {\bf g}_{\rm B}\right) , \label{b_2m^d} \\
&& b_{2m}^{({\rm r})} \equiv  -\frac{R^3}{\displaystyle 1 + \frac{5b}{R}} \int d\Omega \, \bPsi_{2m}^* \cdot \left[\left(\frac{b_{\rm B}}{D_{\rm B}} - \frac{b_{\rm A}}{D_{\rm A}}\right) (\boldsymbol{\nabla} Z)_R\right] . \label{b_2m^r} 
\eea
Since the coefficients $c_{2m}^{(\xi)}$ are related to the tensor of velocity gradients according to Eqs.~(\ref{c_2m^xi}), the coefficients $b_{2m}^{({\rm v})}$ form a contribution that modify the shear viscosity due to the presence of colloidal particles.  The coefficients $b_{2m}^{({\rm d})}$ are associated with the contribution from simple diffusiophoresis and the coefficients $b_{2m}^{({\rm r})}$ from reaction.  The corresponding velocity and pressure fields will be calculated separately in the following subsections~\ref{app:b_2m^v}, \ref{app:b_2m^d}, and~\ref{app:b_2m^r}.

\subsection{The disturbance velocity and pressure due to the velocity gradients}
\label{app:b_2m^v}

According to Eq.~(\ref{asymptotic-velocity}) with ${\bf v}^{\circ}=0$ and the identities ${\bf n}\cdot\bUps_{2m}=Y_{2m}$, ${\bf n}\cdot\bPsi_{2m}=0$, and ${\bf n}\cdot\bPhi_{2m}=0$ , we have that
\be
\sum_{m=-2}^{+2} c_{2m}^{(\xi)} \, Y_{2m} = \frac{1}{2} \, G_{ij} \, n_i \, n_j = \frac{1}{2} \, G_{ij} \, \frac{r_i \, r_j}{r^2} \, ,
\ee
because ${\bf n}={\bf r}/r$, i.e., $n_i=r_i/r$ in Cartesian coordinates.  Therefore, the disturbance velocity and pressure~(\ref{Dv-Dp-b_2m}) due to the contribution of the coefficients~(\ref{b_2m^v}) can be expressed as
\be
\Delta v_i^{({\rm v})} = A_{jk}^{({\rm v})} \, \frac{r_i \, r_j \, r_k}{r^5}
\qquad\mbox{and}\qquad
\Delta p^{({\rm v})} = 2\eta_0 \, A_{jk}^{({\rm v})} \, \frac{r_j \, r_k}{r^5}
\qquad\mbox{with}\qquad
A_{jk}^{({\rm v})} \equiv - \frac{5}{2} \, \frac{\displaystyle 1 + \frac{2b}{R}}{\displaystyle 1 + \frac{5b}{R}} \, R^3 \, G_{jk} \, .
\label{Dv-Dp-b_2m^v}
\ee
Since the tensor of velocity gradients is traceless because of the incompressibility condition, we have that ${\rm tr}\, \boldsymbol{\mathsf A}^{({\rm v})}=A_{jj}^{({\rm v})}=0$.  This contribution goes as the tensor of velocity gradients and provides the well-known correction to the shear viscosity $\eta_0$ of the fluid due to the presence of particles in the suspension \cite{E1906,LP08,PW22}.

\subsection{The disturbance velocity and pressure due to simple diffusiophoresis}
\label{app:b_2m^d}

Since the diffusiophoretic constants of the molecular species may take different values on the two hemispheres according to Eq.~(\ref{diffusio-coeff}), the coefficients~(\ref{X_2m^gd}) are given by
\be
X_{2m}^{(g,{\rm d})} \equiv \frac{1}{4} \sum_{\varkappa={\rm A},{\rm B}} \sum_{h={\rm c},{\rm n}} b_\varkappa^h \, \int d\Omega \, H^h(\theta) \, \bPsi_{2m}^* \cdot {\bf g}_\varkappa \, . \label{X_2m^gd-H} 
\ee
Since the two hemispheres are mapped onto each other by space inversion $(r,\theta,\varphi)\to(r,\pi-\theta,\varphi+\pi)$ and the spherical harmonics satisfy $Y_{lm}(\pi-\theta,\varphi+\pi)=(-1)^l \, Y_{lm}(\theta,\varphi)$, the vectorial spherical harmonics~(\ref{VSH-Psi}) change by $\bPsi_{lm}(\pi-\theta,\varphi+\pi)=(-1)^{l+1} \, \bPsi_{lm}(\theta,\varphi)$.  Therefore,  $\bPsi_{2m}$ is odd under space inversion and Eq.~(\ref{X_2m^gd-H}) becomes
\be
X_{2m}^{(g,{\rm d})} \equiv \frac{1}{4} \sum_{\varkappa={\rm A},{\rm B}} \Delta b_\varkappa \, \int d\Omega \, H^{\rm c}(\theta) \, \bPsi_{2m}^* \cdot {\bf g}_\varkappa
\qquad\mbox{with}\qquad
\Delta b_\varkappa \equiv b_\varkappa^{\rm c} -  b_\varkappa^{\rm n} \, . \label{X_2m^gd-H-i} 
\ee
The calculation of these coefficients with {\tt Mathematica}~\cite{W88} shows that
\bea
&& X_{20}^{(g,{\rm d})} = \frac{3\sqrt{5\pi}}{16} \sum_\varkappa \Delta b_\varkappa \, g_{\varkappa z} \, , \\
&& X_{21}^{(g,{\rm d})} = -X_{2,-1}^{(g,{\rm d})*} = \frac{1}{16}\sqrt{\frac{15 \pi}{2}} \sum_\varkappa \Delta b_\varkappa \left( -g_{\varkappa x} + \imath \, g_{\varkappa y}\right) , \\
&& X_{22}^{(g,{\rm d})} = X_{2,-2}^{(g,{\rm d})*} = 0 \, .
\eea

Now, we can use the following expressions for the spherical harmonics of order two with $m=+1,0,-1$,
\be
Y_{20} = \frac{1}{4}\sqrt{\frac{5}{\pi}} \, \frac{3z^2-r^2}{r^2} \, , \qquad
Y_{21} =  -Y_{2,-1}^* = - \frac{1}{2}\sqrt{\frac{15}{2\pi}} \, \frac{(x+\imath\, y)\, z}{r^2} \, , 
\label{Y_2m-xyz}
\ee
and the assumption that the Janus particle is oriented in the direction ${\bf u}={\bf 1}_z$ to obtain
\be
\sum_{m=-2}^{+2} X_{2m}^{(g,{\rm d})} \, Y_{2m} = \frac{15}{64} \sum_\varkappa \Delta b_\varkappa \left\{ ({\bf g}_\varkappa \cdot{\bf u})\left[ 3 ({\bf u}\cdot{\bf n})^2-1\right] + 2 \left[{\bf g}_\varkappa \cdot {\bf n} - ({\bf g}_\varkappa \cdot{\bf u})({\bf u}\cdot{\bf n})\right] ({\bf u}\cdot{\bf n})\right\} .
\ee
Since ${\bf n}={\bf r}/r=(r_i/r)_{i=x,y,z}$, the disturbance velocity and pressure due to simple diffusiophoresis are given by
\be
\Delta v_i^{({\rm d})} =  A_{jk}^{({\rm d})} \, \frac{r_i \, r_j \, r_k}{r^5} + B^{({\rm d})} \, \frac{r_i}{r^3}
\qquad\mbox{and}\qquad
\Delta p^{({\rm d})} = 2\eta_0 \left( A_{jk}^{({\rm d})} \, \frac{r_j \, r_k}{r^5} + \frac{B^{({\rm d})}}{r^3}\right)
\label{Dv-Dp-b_2m^d}
\ee
with
\bea
&& A_{jk}^{({\rm d})} \equiv - \frac{3\, R^2}{\displaystyle 1 + \frac{5b}{R}} \, \frac{15}{64} \sum_\varkappa \Delta b_\varkappa \left( 2\, g_{\varkappa j} + \, {\bf g}_\varkappa \cdot{\bf u}\, u_j \right) u_k \, , \label{A^d} \\
&& B^{({\rm d})} \equiv \frac{3\, R^2}{\displaystyle 1 + \frac{5b}{R}} \, \frac{15}{64} \sum_\varkappa \Delta b_\varkappa \left({\bf g}_\varkappa \cdot{\bf u}\right) . \label{B^d}
\eea
We note that ${\rm tr}\, \boldsymbol{\mathsf A}^{({\rm d})}=A_{jj}^{({\rm d})}=-3 B^{({\rm d})}$, because $\bf u$ is a unit vector, so that 
\be
\Delta v_i^{({\rm d})} =  A_{jk}^{({\rm d})} \left(\frac{r_j \, r_k}{r^5} - \frac{\delta_{jk}}{3\, r^3}\right) r_i
\qquad\mbox{and}\qquad
\Delta p^{({\rm d})} = 2\eta_0 \, A_{jk}^{({\rm d})} \left(\frac{r_j \, r_k}{r^5} - \frac{\delta_{jk}}{3 \, r^3}\right) ,
\label{Dv-Dp-b_2m^d-2}
\ee
which has the form of a so-called stresslet \cite{B70}.

\subsection{The disturbance velocity and pressure due to reaction}
\label{app:b_2m^r}

Again, the diffusiophoretic constants taking different values on the two hemispheres according to Eq.~(\ref{diffusio-coeff}), we can write
\be
\frac{b_{\rm B}}{D_{\rm B}} - \frac{b_{\rm A}}{D_{\rm A}} = \sum_{h={\rm c},{\rm n}} \Lambda^h \, H^h(\theta) 
\qquad\mbox{with}\qquad
\Lambda^h \equiv \frac{b_{\rm B}^h}{D_{\rm B}} - \frac{b_{\rm A}^h}{D_{\rm A}} \, .
\label{Lambda^h}
\ee
Therefore, the coefficients~(\ref{X_2m^gr}) can be calculated with
 \be
X_{2m}^{(g,{\rm r})}  \equiv \frac{R}{6} \sum_{h={\rm c},{\rm n}} \Lambda^h \int d\Omega \, H^h(\theta) \, \bPsi_{2m}^* \cdot (\boldsymbol{\nabla} Z)_R\, .
\label{X_2m^gr-H}
\ee
Since the function $Z$ has the expansion~(\ref{Z-expansion}), its gradient is given by
\be
R\left(\boldsymbol{\nabla} Z\right)_R = - \sum_{lm} (l+1) \, z_{lm} \, \bUps_{lm} + \sum_{lm} z_{lm} \, \bPsi_{lm} \, .
\ee
We thus have that
\be
R\, \bPsi_{2m}^* \cdot (\boldsymbol{\nabla} Z)_R = \sum_{l'm'} z_{l'm'} \bPsi_{2m}^* \cdot \bPsi_{l'm'}
= \sum_{l'm'} z_{l'm'} \left( \partial_\theta Y_{2m}^* \, \partial_\theta Y_{l'm'} + \frac{1}{\sin^2\theta} \, \partial_\varphi Y_{2m}^* \, \partial_\varphi Y_{l'm'}\right) , 
\ee
because of $\bPsi_{2m}^* \cdot\bUps_{l'm'}=0$ and Eq.~(\ref{VSH-Psi-2}).
Using the property $Y_{lm}(\pi-\theta,\varphi+\pi)=(-1)^l \, Y_{lm}(\theta,\varphi)$ for the spherical harmonics, we find that the coefficients can be expressed as
 \be
X_{2m}^{(g,{\rm r})} = \sum_{l} z_{lm} \left[ \Lambda^{\rm c} + (-1)^l \, \Lambda^{\rm n}\right] {\cal I}_{lm}
\label{X_2m^gr-H-I}
\ee
with the integrals
\be
{\cal I}_{lm} \equiv \frac{1}{6} \int d\Omega \, H^{\rm c}(\theta) \left( \partial_\theta Y_{2m} ^*\, \partial_\theta Y_{lm} + \frac{1}{\sin^2\theta} \, \partial_\varphi Y_{2m}^* \, \partial_\varphi Y_{lm}\right) .
\label{I_lm}
\ee
These integrals can be evaluated using {\tt Mathematica}~\cite{W88}, leading to
\bea
&& X_{20}^{(g,{\rm r})} = \frac{\Lambda^{\rm c}+\Lambda^{\rm n}}{2} \, z_{20} + \left(\Lambda^{\rm c}-\Lambda^{\rm n}\right) \left( \frac{\sqrt{15}}{16} \, z_{10} + \frac{\sqrt{35}}{16} \, z_{30} - \frac{5\sqrt{55}}{256} \, z_{50} + \frac{7\sqrt{3}}{128} \, z_{70} - \frac{15\sqrt{95}}{2048} \, z_{90} + \cdots \right) , \label{X_20^gr-z_lm}\\
&& X_{21}^{(g,{\rm r})} = \frac{\Lambda^{\rm c}+\Lambda^{\rm n}}{2} \, z_{21} + \left(\Lambda^{\rm c}-\Lambda^{\rm n}\right) \left( \frac{\sqrt{5}}{16} \, z_{11} + \frac{1}{8}\sqrt{\frac{35}{2}} \, z_{31} - \frac{25\sqrt{11}}{256} \, z_{51} + \frac{7\sqrt{7}}{64} \, z_{71} - \frac{75\sqrt{57}}{2048} \, z_{91} + \cdots \right) , \qquad\label{X_21^gr-z_lm}\\
&& X_{2,-1}^{(g,{\rm r})} = - X_{21}^{(g,{\rm r})*} \, , \qquad X_{22}^{(g,{\rm r})} = X_{2,-2}^{(g,{\rm r})*} = 0 \, , \label{X_21^gr-X_22^gr-z_lm}
\eea
where the last line results from the facts that $z_{l,-1}=-z_{l1}^*$ and $z_{l,\pm 2}=0$ according to Eqs.~(\ref{z_l1}) and~(\ref{z_lm}).  Now, using the values of the coefficients $z_{lm}$ given by Eqs.~(\ref{z_l0})-(\ref{z_lm})  in terms of the coefficients~(\ref{a_l-b_l-c_l}) with Eqs.~(\ref{A_l-B_l-C_l}) and~(\ref{M-N}), we find that
\bea
&& X_{20}^{(g,{\rm r})} = \sqrt{\frac{4\pi}{5}} \left( \alpha_\Lambda \, \varrho + \frac{3R}{2} \, \beta_\Lambda \, g_{\varrho z} \right) , \label{X_20^gr-alpha-beta}\\
&& X_{21}^{(g,{\rm r})} = - X_{2,-1}^{(g,{\rm r})*} =  \sqrt{\frac{2\pi}{15}} \frac{3R}{2} \, \gamma_\Lambda \left(- g_{\varrho x} + \imath \, g_{\varrho y}\right) , \label{X_21^gr-gamma}\\
&& X_{22}^{(g,{\rm r})} = X_{2,-2}^{(g,{\rm r})*} = 0 \, , \label{X_22^gr-z_lm}
\eea
where
\bea
&& \alpha_\Lambda \equiv \frac{\Lambda^{\rm c}+\Lambda^{\rm n}}{2} \, a_2 + \left(\Lambda^{\rm c}-\Lambda^{\rm n}\right) \left( \frac{5}{16} \, a_1 + \frac{5}{16} \, a_3 - \frac{25}{256} \, a_5 + \frac{7}{128} \, a_7 - \frac{75}{2048} \, a_9 + \cdots \right) ,\label{alpha_Lambda} \\
&& \beta_\Lambda \equiv \frac{\Lambda^{\rm c}+\Lambda^{\rm n}}{2} \, b_2 + \left(\Lambda^{\rm c}-\Lambda^{\rm n}\right) \left( \frac{5}{16} \, b_1 + \frac{5}{16} \, b_3 - \frac{25}{256} \, b_5 + \frac{7}{128} \, b_7 - \frac{75}{2048} \, b_9 + \cdots \right) , \label{beta_Lambda} \\
&& \gamma_\Lambda \equiv \frac{\Lambda^{\rm c}+\Lambda^{\rm n}}{2} \, c_2 + \left(\Lambda^{\rm c}-\Lambda^{\rm n}\right) \left( \frac{5}{16} \, c_1 + \frac{5}{16} \, c_3 - \frac{25}{256} \, c_5 + \frac{7}{128} \, c_7 - \frac{75}{2048} \, c_9 + \cdots \right) . \label{gamma_Lambda}
\eea

Using Eq.~(\ref{Y_2m-xyz}) and the assumption ${\bf u}={\bf 1}_z$, we find that
\be
\sum_{m=-2}^{+2} X_{2m}^{(g,{\rm r})} \, Y_{2m} = \frac{1}{2} \left( \alpha_\Lambda \, \varrho + \frac{3R}{2} \, \beta_\Lambda \, {\bf g}_{\varrho}\cdot{\bf u} \right)\left[ 3 ({\bf u}\cdot{\bf n})^2-1\right] +  \frac{3R}{2} \, \gamma_\Lambda \left[{\bf g}_{\varrho}\cdot{\bf n}-({\bf g}_{\varrho}\cdot{\bf u})({\bf u}\cdot{\bf n})\right] ({\bf u}\cdot{\bf n})\, .
\ee
Consequently, the disturbance velocity and pressure due to reaction are given by
\be
\Delta v_i^{({\rm r})} = A_{jk}^{({\rm r})} \, \frac{r_i \, r_j \, r_k}{r^5} + B^{({\rm r})} \, \frac{r_i}{r^3}
\qquad\mbox{and}\qquad
\Delta p^{({\rm r})} = 2\eta_0 \left( A_{jk}^{({\rm r})} \, \frac{r_j \, r_k}{r^5} + \frac{B^{({\rm r})}}{r^3} \right)
\label{Dv-Dp-b_2m^r}
\ee
with
\bea
&& A_{jk}^{({\rm r})} \equiv - \frac{3\, R^2}{\displaystyle 1 + \frac{5b}{R}} \, \frac{3}{2} \left[\left( \alpha_\Lambda \, \varrho + \frac{3R}{2} \, \beta_\Lambda \, {\bf g}_{\varrho}\cdot{\bf u}\right) u_j + R \, \gamma_\Lambda \left( g_{\varrho j} - {\bf g}_{\varrho}\cdot{\bf u} \, u_j \right) \right] u_k \, , \label{A^r}\\
&& B^{({\rm r})} \equiv \frac{3\, R^2}{\displaystyle 1 + \frac{5b}{R}} \, \frac{1}{2} \left( \alpha_\Lambda \, \varrho + \frac{3R}{2} \, \beta_\Lambda \, {\bf g}_{\varrho}\cdot{\bf u} \right) . \label{B^r}
\eea
Here also, we have the property that ${\rm tr}\, \boldsymbol{\mathsf A}^{({\rm r})}=A_{jj}^{({\rm r})}=-3 B^{({\rm r})}$, because $\bf u$ is a unit vector, so that this contribution is given by the following stresslet,
\be
\Delta v_i^{({\rm r})} = A_{jk}^{({\rm r})} \left(\frac{r_j \, r_k}{r^5} - \frac{\delta_{jk}}{3 \, r^3} \right) r_i 
\qquad\mbox{and}\qquad
\Delta p^{({\rm r})} = 2\eta_0 \, A_{jk}^{({\rm r})} \left(\frac{r_j \, r_k}{r^5} - \frac{\delta_{jk}}{3 \, r^3} \right) .
\label{Dv-Dp-b_2m^r-2}
\ee

\section{The calculation of the change in the stress tensor}
\label{app:P_ij}

Here, we calculate the change~(\ref{DPij-suspension-LL}) of the stress tensor due to the disturbance velocity and pressure~(\ref{Dv-Dp-stresslet}).  The latter can be equivalently written as
\be
\Delta v_i = \Xi \, r_i
\qquad\mbox{and}\qquad
\Delta p = 2\eta_0 \, \Xi
\qquad\mbox{with}\qquad
\Xi \equiv A_{jk} \left(\frac{r_j \, r_k}{r^5} - \frac{\delta_{jk}}{3 \, r^3} \right) ,
\label{Xi-dfn}
\ee
up to terms having higher powers of $r^{-1}$ and leading to negligible corrections in the limit $r\to\infty$, i.e., in the dilute-system limit.
Inserting the expressions~(\ref{Xi-dfn}) into Eq.~(\ref{DPij-suspension-LL}), we get
\be
\Delta\sigma_{ij} = \eta_0 \int d^2u\, f_{\rm C} \int \left( -\Xi  \, r_j \, d\Sigma_i - \Xi \, r_i \, d\Sigma_j + r_i \, r_j \, \nabla_k\Xi \, d\Sigma_k + r_j \, r_k \, \nabla_i\Xi \, d\Sigma_k \right) .
\label{DPij-suspension-LL-2}
\ee
The derivatives $\nabla_i\Xi=\partial\Xi/\partial r_i$ of the function $\Xi$ can be calculated.  Afterwards, the function $\Xi$ and its derivatives can be expressed in terms of the unit vector $n_i=r_i/r$, giving
\be
\Xi = \frac{A_{jk}}{r^3} \left(n_j \, n_k- \frac{1}{3}\, \delta_{jk} \right) , \qquad
\nabla_i\Xi = \frac{A_{jk}}{r^4} \left(\delta_{ij} \, n_k + \delta_{ik} \, n_j + \delta_{jk} \, n_i - 5 \, n_i \, n_j \, n_k \right) ,
\ee
and a similar expression for $\nabla_k\Xi$.  Moreover, we have that $r_i = r \, n_i$ and $d\Sigma_i = r^2 \, n_i \, d\Omega$.  Therefore, Eq.~(\ref{DPij-suspension-LL-2}) becomes
\be
\Delta\sigma_{ij} = 4\pi\eta_0 \int d^2u\, f_{\rm C} \, A_{kl} \left( -10 \, \overline{n_i n_j n_k n_l} + \frac{8}{3} \, \overline{n_i n_j}  \, \delta_{kl} + \overline{n_j n_l}  \, \delta_{ik} + \overline{n_j n_k}  \, \delta_{il} \right)
\label{DPij-suspension-LL-3}
\ee
with the notation $\overline{(\cdot)} \equiv (4\pi)^{-1} \int (\cdot) \, d\Omega$. Since
\be
\overline{n_i n_j} = \frac{1}{3} \, \delta_{ij}
\qquad\mbox{and}\qquad
\overline{n_i n_j n_k n_l} = \frac{1}{15} \left( \delta_{ij} \, \delta_{kl} +  \delta_{ik} \, \delta_{jl} +  \delta_{il} \, \delta_{jk}\right) ,
\ee
we finally obtain the result given by Eq.~(\ref{P_ij-A_ij}).

\section{The particle velocities in the case of uniform diffusiophoretic constants}
\label{app:velocities-uniform}

The force-free and torque-free conditions fully determine the translational and rotational velocities of the Janus particles in the active suspension.  Indeed, using the force-free condition ${\bf F}=0$ with Eq.~(\ref{force-single}) and the torque-free condition ${\bf T}=0$ with Eq.~(\ref{torque-single}), we obtain the following expressions for the particle velocities,
\bea
{\bf V} &=& {\bf v}^{\circ} + \frac{1}{\displaystyle 1 +\frac{2b}{R}}  \sum_\varkappa \overline{b_\varkappa \left({\boldsymbol{\mathsf 1}}-{\bf n}{\bf n}\right) \cdot \boldsymbol{\nabla}c_\varkappa}^{\rm s} \, ,  \label{transl-velocity-single} \\
\boldsymbol{\Omega} &=& \frac{1}{2}\, \boldsymbol{\omega}_\infty + \frac{3}{2 R^2}  \sum_\varkappa \overline{b_\varkappa {\bf r}\times \boldsymbol{\nabla}c_\varkappa}^{\rm s} \, .
\label{rot-velocity-single}
\eea
These velocities have been calculated in Ref.~\cite{GK20} for diffusiophoretic constants taking different values on the two hemispheres according to Eq.~(\ref{diffusio-coeff}).

In the case $b_\varkappa^{\rm c}=b_\varkappa^{\rm n}$, where the diffusiophoretic constants take uniform values, the general expressions obtained in Ref.~\cite{GK20} can be simplified using the following relations between the quantities defined therein and those used in the present article,
\be
\alpha^{\rm c} + \alpha^{\rm n} = -\frac{4}{3}\, a_1 \, , \qquad
\beta^{\rm c} + \beta^{\rm n} = -2 \, b_1 \, , \qquad
\gamma^{\rm c} + \gamma^{\rm n} = -2\, c_1 \, , \qquad
\delta^{\rm c} + \delta^{\rm n} = 0 \, , 
\qquad\mbox{and}\qquad
Y^{\rm c} = Y^{\rm n} = - \Lambda \, ,
\ee
where the coefficients $a_1$, $b_1$, and $c_1$ are given by Eqs.~(\ref{a_l-b_l-c_l}), (\ref{A_l-B_l-C_l}), and~(\ref{M-N}), and $\Lambda$ is introduced in Eq.~(\ref{uniform-Lambda}).
Accordingly, the translational velocity of a Janus particle can be expressed as
\be
{\bf V} = {\bf v}^{\circ} + \frac{1}{\displaystyle 1 +\frac{2b}{R}}  \left[ b_{\rm A} \, {\bf g}_{\rm A} +  b_{\rm B} \, {\bf g}_{\rm B} + \frac{2}{3} \, a_1 \Lambda\, \varrho \, {\bf u} + R \, c_1 \, \Lambda \, {\bf g}_\varrho + R \, (b_1 - c_1) \, \Lambda \, {\bf g}_\varrho \cdot{\bf u} \, {\bf u} \right]
\label{V-single-particle}
\ee
with ${\bf v}^{\circ}=\overline{\bf v}$ \cite{GK20}. In the absence of molecular concentration gradients, i.e., if ${\bf g}_{\rm A}={\bf g}_{\rm B}=0$, the vectorial particle velocity~(\ref{V-single-particle}) becomes ${\bf V} = {\bf v}^{\circ} + V^{({\rm r})} \, {\bf u}$, as written in terms of the scalar particle velocity~(\ref{V^r}), as previously shown in Refs.~\cite{GK18,GK19}.

Furthermore, in the case of uniform diffusiophoretic constants, the rotational velocity~(\ref{rot-velocity-single}) of a Janus particle reduces to
\be
\boldsymbol{\Omega} = \frac{1}{2}\, \boldsymbol{\omega}_\infty = \frac{1}{2}\, \overline{\boldsymbol{\nabla}\times{\bf v}}
\ee
with a contribution from diffusiophoresis equal to zero \cite{GK18,GK19,GK20}.


\end{document}